\documentclass[conference]{IEEEtran}
\usepackage[sort,compress]{cite}
\usepackage{amsmath,amssymb,amsfonts}
\usepackage{graphicx}
\usepackage{booktabs}
\usepackage{upgreek}

\newcommand{\ra}[1]{\renewcommand{\arraystretch}{#1}}
\usepackage{xspace}

\begin{document}
\graphicspath{{./figures/}}
\title{A Review of Sc-containing ``Scandate" \\Thermionic Cathodes
}

\author{\IEEEauthorblockN{Mujan N. Seif, Qunfei Zhou, Xiaotao Liu, T. John Balk, Matthew J. Beck}
\IEEEauthorblockA{\textit{Department of Chemical and Materials Engineering}}
\textit{University of Kentucky}\\
Lexington, KY, USA\\
m.beck@uky.edu}

\maketitle

\begin{abstract}
Although thermionic emission has been studied for more than 100 years, recent interest in novel electron devices for military and civilian use has led to a surge in demand for cathodes with enhanced emission properties (e.g. higher current density, more uniform emission, lower operating temperatures, or extended in-service longevity). 
Sc-containing ``scandate" cathodes have been widely reported to exhibit superior emission properties compared to previous-generation thermionic cathodes, including oxide, B-, and M-type cathodes. 
Despite extensive study spanning several decades, the mechanism by which the addition of Sc enhances cathode emission remains ambiguous, and certain limitations -- non-uniform emission, low reproducibility, inconsistent longevity -- continue to prevent widespread commercial integration of scandate cathodes into electron devices. 
This review attempts to survey the literature to-date addressing the fabrication, structure, and properties of scandate cathodes, with particular attention to studies addressing the role of Sc in enhancing emission.
\end{abstract}


\section{Introduction}
As long-life, high current density, and high total current electron sources, thermionic cathodes are essential components in a wide array of vacuum electron devices (VEDs) utilized in both military and civilian applications.
Key applications include traveling wave tubes in satellite or long-range communication systems, traveling wave linear accelerators, magnetrons in microwave ovens and RADAR systems, electron sources for nuclear magnetic resonance, electron microscopes, e-beam lithography systems, and more \cite{Wilson:PI:2007,Zavadil:ACP:2001,King:3IECECE:2000,Jiang:I25IVESCPICN:2004,Green:2IIVEC:2008,Broers:APL:1973,Gubiotti:ALTV:2013,Gaertner:JoVS&TB:2012}.
Increasingly, high power, small footprint cathodes are being sought for terahertz or mm-wave devices operating beyond the crowded kHz, MHz, and GHz frequency ranges \cite{Pi:ICM:2011}. 
In response to these demands for high current density emitters and novel electronic devices, Sc-containing ``scandate'' cathodes offering order-of-magnitude increases in emission current density and significant decreases in required operating temperatures have received a great deal of attention. 
Multiple investigations \cite{Vancil:ITED:2014,Wang:ITED:2009,Lesny:ITED:1990,Raju:ITED:1994,Zhao:ITED:2011,Yuan:ASS:2005,Hasker:ASS:1986,Sasaki:ASS:1997} have demonstrated the capabilities of scandate cathodes.
Nevertheless, outstanding issues associated with both the reliable fabrication and operation of scandate cathodes---including non-uniform emission, low repeatability, and unpredictable longevity---continue to limit their application \cite{Vancil:ITED:2014,Crombeen:ITED:1990,Liu:I25IVESCPICN:2004,Gibson:ITED:1989,Sasaki:ASS:1999,Shih:ASS:2002,Gilmour:D:1986,Jennison:SS:2004}.
\begin{table*}
	\centering
	\ra{1.3}
	\label{Table: summary of properties}
	\caption{Summary of thermionic cathode properties.}
	\vspace{-1.5 mm}
	\begin{tabular}{ccccccc} \toprule
		Type & Matrix & Impregnant & Coating & Emission (A/cm$^2$) & Operating Temperature ($^\circ$C) & $\Phi$ (eV) \\
		\hline
		Oxide        & Ni \cite{Madjid:PRL:1972,Jenkins:V:1969,Yamamoto:ASS:1984} & - & Ba-Sr-O \cite{Jenkins:V:1969,Yamamoto:RPP:2006} & 1.0 \cite{Jenkins:V:1969,Yamamoto:RPP:2006}               & 700-800\cite{Jenkins:V:1969,Yamamoto:RPP:2006}                      & 1.5\cite{Jenkins:V:1969,Yamamoto:RPP:2006}      \\
		B-type       & W & Ba-Ca-Al-O* & - & 5.0\cite{Ives:ITPS:2010,Rittner:JAP:1977}                 & 1150\cite{Muller:ITED:1989}                      & 2.0-2.1\cite{Yamamoto:RPP:2006,Muller:ITED:1989}                    \\
		M-type       & W & Ba-Ca-Al-O* & Os-Ru \cite{Cronin:IP:1981}, Os \cite{Zalm:PTR:1966},  & 1-100\cite{Zalm:PTR:1966,Chopra:IJoR:1994,Li:ASS:2005,Ives:ITPS:2010,Shih:ITED:1987,Aida:JAP:1993}              &  950-1050\cite{Zalm:PTR:1966,Li:ASS:2005,Aida:JAP:1993}                      &   1.16-1.72\cite{Swartzentruber:JoVS&TA:2014,Zalm:PTR:1966}               \\
		 & & & Ru\cite{Bhide:JPDAP:1970}, Ir \cite{Bhide:JPDAP:1970} & & & \\
		Scandate     & W, W+Sc &  Sc/Ba-Ca-Al-O* & ** &  1-400 \cite{Gartner:ASS:1997,Yuan:ASS:2005}   & 750-950 \cite{Kirkwood:ITED:2018} & $\sim$1.15-1.5 \cite{Gartner:ASS:1997,Kirkwood:ITED:2018}   \\
		\bottomrule 
	\end{tabular}
	\begin{tabular}{lllllllllllllllllllllllllllllllllllllllllllllllllllll}
	 * This is an oxide mixture containing Ba, Ca, and Al (and Sc in scandate cathodes).     &&&&&&&&&&&&&&&&&&&&&&&&&&&&&&&&&&&&&&&&&&&&&&&&&&&&\\
	 ** Top-layer scandate cathodes have a W+Sc$_2$O$_3$ coating \cite{Hasker:ASS:1986,Gartner:ASS:1997}.
	 &&&&&&&&&&&&&&&&&&&&&&&&&&&&&&&&&&&&&&&&&&&&&&&&&&&&
	\end{tabular}
\end{table*}
Scandate cathodes represent the most modern evolution of dispenser cathodes and a continuation of long-standing efforts to increase emitted current density while lowering operating temperature and increasing cathode life and reliability.
While thermionic cathodes have been studied since the time of Edison, ``dispenser'' cathodes were first developed in the 1950s.
Previous ``oxide'' cathodes consisted of a solid metal core coated in an emission enhancing oxide layer (Ba, Sr, or Ca carbonates and/or oxides, or other mixed oxides)\cite{Herrmann::1951,Beck:PIBECE:1959,Beck:N:1954}.
Oxide cathodes exhibited effective work functions as low as $\sim$1.5 eV \cite{Herrmann::1951,Jenkins:V:1969,Yamamoto:RPP:2006} and emitted between 1-10 A/cm$^2$ at operating temperatures of 700-800$^\circ$C but were limited to low total currents because emission was constrained by the conductivity of the oxide layer \cite{Jenkins:V:1969,Yamamoto:RPP:2006}. 
The first dispenser cathodes (``L-types''), by contrast, consisted of a porous refractory metal pellet atop a reservoir of emission enhancing oxides.
During operation at high temperature, these oxide-containing reservoirs ``dispensed'' Ba to the emitting surface of the cathode, resulting in reduced effective work functions and enhanced emission \cite{Lemmens:PTR:1950,Rittner:JAP:1957,Rutledge:JAP:1957,Gaertner:ASS:2005}.

In 1951, a cathode consisting of a pressed pellet of mixed Ba aluminate and W powder was developed that greatly improved performance compared to L-type cathodes.
Researchers at Philips Laboratories demonstrated that performance was further improved by impregnating a previously pressed pure W pellet with Ba aluminate (rather than pressing mixed powders directly) \cite{Levi::1955}.
The further inclusion of CaO in the impregnant was reported to reduce the evaporation rate of Ba during operation and to greatly enhance cathode emission properties \cite{Levi:JAP:1955}.
Cathodes of this type, formed by first pressing a porous pellet from W powder then impregnating the pores with a mixture of BaO, CaO, and Al$_2$O$_3$, were called impregnated (and later ``B-type'') cathodes.
B-type cathodes exhibit work functions of $\sim$2 eV and yield emitted current densities of $\sim$1 A/cm$^2$ at 1150$^\circ$C \cite{Norman:PRL:1987,Tuck:V:1983}. 

Around 1964, it was found that coating a B-type cathode with certain refractory metals (e.g. Os, Ir, or Ru) markedly lowered the cathode's effective work function \cite{Cronin:IP:1981,Li:ASS:2005,Green:AoSS:1981,Barik:ASS:2013}. 
This type of cathode was termed the ``M-type'' (magic) cathode. 
Although the work functions of the refractory metal coatings are higher than that of W itself, the overall effective work function of the emitting surface---understood to consist of an atomically thin layer of BaO atop the refractory metals coating the W grains---is lower than that of B-type cathodes. 
M-type cathodes exhibit effective work functions of 1.1--1.8 eV and produce emitted current densities on the order of 10 A/cm$^2$ at about 1050$^\circ$C \cite{Zalm:PTR:1966,VanStratum::}.

Around the same time, Figner et al. \cite{Figner::1967} introduced one of the first Sc-containing cathodes, fabricated by incorporating Ba scandate into a porous refractory metal matrix. 
The electron emission behavior of this cathode was comparable to B- and M-type cathodes but was reported to remain stable in air for longer periods and did not require a preliminary heating stage prior to operation.
These Sc-containing cathodes were similar to B-type cathodes---porous W pellets that have been impregnated by a BaO/CaO/Al$_2$O$_3$ mixture---but distinguished by the inclusion of Sc$_2$O$_3$ either in the impregnant, or as part of the W pellet. 
Scandate cathodes \cite{Figner::1967,Deckers::1997,Deckers::1999,Snijkers::1993} have reported effective work functions of $\sim$1.15 eV and emitted current densities between 100-400 A/cm$^2$ at operating temperatures of only $\sim$850$^\circ$C.
Numerous studies support the notion that scandate cathodes significantly outperform all previous thermionic cathodes, particularly on the basis of emitted current density \cite{Gartner:ASS:1997,Gartner:ASS:1999,Wang:ITED:2015,Yamamoto:RPP:2006}.

Despite more than fifty years of scandate cathode research, there is still widespread debate as to the mechanism by which the addition of Sc affects cathode performance---or even whether it is Sc or Sc$_2$O$_3$ that drives the effect.
This is at least a partial explanation for why researchers have still not demonstrated sufficient understanding of scandate cathode fabrication and operating conditions to overcome variations in reliability, operating performance, and cathode lifetime that limit integration of these cathodes into most VED applications.
Extensive reviews published before the 1960s \cite{Richardson::1921,Dushman:RMP:1930,Nottingham:HdP:1956,Becker:RMP:1935,Herring:RMP:1949} focused on the fundamental physics of thermionic electron emission, and the behavior of early-stage thermionic cathodes (primarily oxide cathodes). 
Subsequent reviews \cite{Yamamoto:RPP:2006,Jenkins:V:1969,Cronin:IP:1981,Tuck:V:1983,Green:2IIVEC:2008} focused on the evolution, fabrication, and operation of various classes of thermionic cathodes, including some discussion of scandate cathodes. 
Here we present a comprehensive review of efforts to understand and optimize scandate cathode fabrication and performance and explore basic research perspectives regarding the physical origins of scandate cathodes' exceptional emission characteristics. 

\section{Background}
\subsection{Thermionic Emission and the Work Function}
The physics underpinning thermionic emission itself have been detailed in a number of comprehensive and insightful texts. 
O.W. Richardson \cite{Richardson::1929}, who won the Nobel Prize in Physics in 1928 for his contributions relating to thermionic emission, used classical free electron theory to write an expression for thermionic electron emission current as a function of temperature \cite{Richardson::1921}.
Richardson's theory was then appended by Dushman \cite{Dushman:RMP:1930}, who used quantum theory to assign physical meaning to the constants in the Richardson-Dushman equation (Eq.\ref{richardson_eq}):
\begin{equation}
J = A_\textrm{R}T^2e^{\frac{-\Phi}{\textrm{k}_\textrm{B}T}} \label{richardson_eq}
\end{equation}
Here, $J$ is the emitted electron flux (current density), and $T$ the temperature of the emitting material. 
The work function $\Phi$ characterizes the energy barrier for electron escape from a surface, and $\textrm{k$_\textrm{B}$}$ is the Boltzmann constant. 
The Richardson constant, $A_\textrm{R}$, accounts for the number of electrons available for emission and is $A_\textrm{R}$ = exp(4$\pi$m$_e$k$^2$/h$^2$) = 120 A/cm$^2$K$^2$ when the electron source is an ideal free-electron gas \cite{Dushman:RMP:1930,Nottingham:HdP:1956}. 
For real materials and surfaces, $A_\textrm{R}$ varies strongly and is determined by details of a material's electronic structure \cite{Eftekhari:PSSA:1993,Missous:JoAP:1991,Srivastava:SE:1981}.

Thorough summaries of the physics underlying thermionic emission from both classical and quantum mechanical perspectives have been provided by Nottingham \cite{Nottingham:HdP:1956} and Murphy and Good \cite{Murphy:PR:1956} and include discussions of the effects of temperature, Schottky image potential, and space-charge effects on emission from uniform surfaces. 
More recent reports have extended and updated our understanding of the physics of thermionic emission and have demonstrated efficient and accurate calculations of emission currents using modern computing techniques \cite{Voss:JCP:2013,Voss:PRA:2014,Jensen:PRST-AB:2014,Chou:JPCM:2012,Chou:JPCC:2014}.

It should be noted that the Richardson-Dushman equation has been used to compare the performance of thermionic cathodes \cite{Yamamoto:RPP:2006} of various and complex structure by allowing assignment of an ``effective'' work function to a cathode based on measurements of the temperature dependence of emitted thermionic current.
While Becker \cite{Becker:RMP:1935} in 1935 considered any distinction between ideal, theoretical work functions characterizing bulk materials and an ``effective'' work function characterizing a cathode device unnecessary, emission from practical cathodes does not arise from pure, homogeneous, flat surfaces that can be defined by a single fundamental work function as envisioned in the Richardson-Dushman theory\cite{Nottingham:HdP:1956,Herring:RMP:1949}.
Instead, cathode emission arises from an aggregation of phases, grains, and facets, and a complex, rough, and non-uniform emitting surface that, when treated as a whole, must be characterized by an effective work function.

\subsection{Effects of surface dipoles}
Adsorption of cations on emitting surfaces also affects emission and effective work function.
The presence of surface cations results in a net surface dipole that lowers the electrostatic potential barrier to electron emission, reducing the work function.
The magnitude of this effect increases with increasing coverage of adsorbed cations to some maximum where increased planar density of cations reduces the effective per cation charge\cite{Ishida:PRB:1988,Ishida:PRB:1990,Soukiassian:PRB:1985,Bonzel:SSR:1988}.
This effect was first harnessed with alkali or alkali-earth metal ions adsorbed on oxide cathode surfaces, and is the key mechanism behind the low work functions and high emission currents realizable from dispenser cathodes based on porous W.

The relationship between work function and dipole density is modeled by the Helmholtz equation (Eq.~\ref{wfdipoledensity}) \cite{Vlahos:PRB:2010,Holzl:SSP:1979,Gaines::1966}:
\begin{equation}
	\Delta\Phi = - \frac{\textbf{e}}{\epsilon_0}\mu_{0}(N)N\label{wfdipoledensity}
\end{equation}
Here $\Delta\Phi$ is the change in work function due to surface adsorption, $\epsilon_0$ is the permittivity of free space, $\mu_{0}(N)$ is the normal component of the dipole moment (assuming the $z$ direction is normal to the surface), and $N$ is the dipole density per unit surface area. 
However, when the dipole density, or coverage of adsorbed atoms is too large, work function was found to increase with coverage.
This was attributed to depolarization that leads to a decrease in dipole moment \cite{Ishida:PRB:1988,Ishida:PRB:1990,Gorodetsky:SS:1977,Muller:ASS:1997,Cortenraad:ASS:2002}. 
As a result, $\Delta\Phi$, considering both polarization and depolarization, was given as Eq. \ref{pole_depole_wf}\cite{Vlahos:PRB:2010,Holzl:SSP:1979}:
\begin{equation}
	\Delta\Phi = -\frac{\textbf{e}N\mu_{0_z}}{\epsilon_0(1+(\textrm{c}\alpha N^{3/2})/(4\pi\epsilon_0)} \label{pole_depole_wf}
\end{equation}
In this expression, $\upmu_{0{_z}}$ is the surface-normal dipole moment in the limit of very low surface coverage.
The variable c is a dimensionless parameter related to the surface coverage of adsorptions. 
For a square and triangular network of dipoles, c can be approximated as 9.0 \cite{Vlahos:PRB:2010, Holzl:SSP:1979, Topping:PotRSoLSACPoaMaPC:1927}. 
The parameter $\alpha$ is the polarizability of the surface dipoles, typically computed using fits that yield c$\alpha$ \cite{Vlahos:PRB:2010}. 
In this classical model, the dipole strength is assumed to be independent of surface geometry, only responding linearly to local electric fields.
Meanwhile, the values of c and $\alpha$ were fitted to match experimental and computational results. 
Thus, direct calculation of work function reduction by Eq.~\ref{pole_depole_wf} is prohibitively difficult.

For the case of $6s$ alkali/alkali-earth metals, the assumption of the electrostatically interacting point-dipoles made in classical models is inoperative due to the large extent of the $6s$ state. 
Significant orbital overlap between adsorbate-substrate and adsorbate-adsorbate, especially at the intermediate region where minimum work function is achieved, was found for Cs-adsorbed W surfaces \cite{Soukiassian:PRB:1985}.
Therefore, an orbital-overlap model beyond the classical description was proposed by Chou et al. \cite{Chou:JPCM:2012}, where the work function reduction was expressed as Eq.~\ref{orboverlapwf}:
\begin{equation}
\Delta\Phi = \frac{\textrm{c}_1N}{{1+\textrm{c}_2N^{3/2}(1-e^{-\textrm{c}_3/N})+\textrm{c}_4e^{-\textrm{c}_3/N}}}  \label{orboverlapwf}
\end{equation}
Here, the variable c$_3$ is inversely proportional to the spatial spread of $6s$ Gaussian states of alkali/alkali-earth metals and c$_4$ represents the strength of depolarization caused by orbital overlap. 
The factor [1-exp(-c$_3$/$N$)] is selected as the simplest form to scale down the diverging classical depolarization term. 
The additional depolarization term c$_4$exp(-c$_3$/$N$) describes the transfer of charge density from the dipoles normal to the surface to in-plane orbitals which result in covalent bonding between adsorbates. 
This model works well, especially at high adsorbate coverage, and results obtained from this model are reported in Ref.~\cite{Chou:JPCC:2014}.

\subsection{B- and M-type Cathodes}\label{B-M-types}
B- and M-type dispenser cathodes are the direct antecedent of modern Sc-containing cathodes and consist of a porous W pellet impregnated with a Ba-containing mix of oxides.
The Ba-containing impregnant is understood to be a source of Ba cations that form a dipole layer at the emitting W surfaces, reducing the cathode's effective work function.
A range of studies on B- and M-type cathodes have reported that Ba cations are present at the emitting surface in the form of a layer of adsorbed Ba-O complexes up to one monolayer thick (as opposed to Ba atoms or bulk-like BaO layers or particles)\cite{Rittner:JAP:1977,Haas:JoAP:1975,Forman:JAP:1976,Jones:ASS:1979}.
Here, ``one monolayer'' represents approximately one Ba-O complex per W surface atom.
The presence of Ba in the form of Ba-O complexes is supported by a number of Auger electron spectroscopy (AES) studies that report electron binding energies for Ba on cathode surfaces that fall between those characteristic of metallic Ba and bulk BaO \cite{Lesny:ITED:1990,Cortenraad:JoAP:2001,Norman:PRL:1987,Haas:ASS:1985,Ibrahim:PRB:2007,Lai:ASS:2018,Lamouri:PRB:1994,Lamouri:SIA:1994,Forman:AoSS:1984}.
Separately, indirect measurements of adsorbate binding energies \cite{Forman:ASS:1979,Cortenraad:JoAP:2001}, as well as DFT-computed binding energies for various potential adsorbate species\cite{Zhou:ASS:2018,Vlahos:PRB:2010,Vlahos:APL:2009,Seif:ASS:2021} both show that Ba-O complexes are the most stable adsorption species on W surfaces, particularly compared to atomic Ba.
Norman et al. \cite{Norman:PRL:1987} used surface extended X-ray adsorption fine structure (SEXAFS) to detect the radial distance and types of neighboring atoms for surface adsorptions from the oscillatory part of the photon absorption coefficient. 
They found that the arrangement of Ba and O was Ba atop O adsorbed on W, with one Ba atom attached to \emph{one} O atom in B-type cathodes and one Ba atom to \emph{two} O atoms in M-type cathodes. 
Complementary computational work conducted by Ibrahim and Lee \cite{Ibrahim:PRB:2007} also reported that the atomic arrangement of Ba on top of O adsorbed on W is more stable than O atop Ba and could yield work function close to those reported for dispenser cathodes.

A handful of studies have specifically examined the coverage of Ba/Ba-O on B- and M-type cathode surfaces.
Utilizing x-ray photoelectron spectroscopy (XPS), van Veen \cite{VanVeen:ASS:1987} found Ba densities on B-type and M-type cathodes of 4.0$\times$10$^{14}$ and 5.8$\times$10$^{14}$ atoms/cm$^2$, respectively, similar to the surface density of atoms in W, approximately 10.0$\times$10$^{14}$ atoms/cm$^2$.
Studies of Ba/O and Ba/W AES peak ratios suggest partial layer coverages of Ba on W \cite{Forman:ASS:1979,Haas:ASS:1985,Cortenraad:ASS:2002}.
Shih et al. \cite{Shih:ASS:2005} hypothesized that Ba layers were formed only in O$_2$-deficient conditions since the species found to evaporate from cathode surfaces at application-relevant conditions were dominated by metallic Ba.
At higher O$_2$ content, temperature-programmed desorption (TPD) and AES spectra showed that, instead of fully covering the W surface, deposited Ba with O preferentially formed BaO clusters. 

\subsection{Emerging Scandate Cathodes}
As part of ongoing efforts to both increase emitted thermionic current density and reduce required operating temperatures, Sc-containing (or ``scandate'') cathodes have evolved over the past 50 years from B- and M-type cathodes.
Like B- and M-cathodes, scandate cathodes are based on a porous W pellet that is impregnated by melting and calcining a mix of Ba, Ca and Al oxides placed atop the cathode.
Unlike B- and M-cathodes, Sc or Sc$_2$O$_3$ is added either in the mixed oxide impregnant material or as a dopant in the W pellet itself.
The weight percent of Sc-containing material added to form a scandate cathode generally amounts to no more than a few (weight) percent of the added impregnant.
Scandate cathodes are also ``activated'' (described in Refs.~\cite{Beck:JEC:1963,Sandor:JEC:1962,VanOostrom:AoSS:1979,Roquais:MDiVES:2020}) like B- and M-cathodes by annealing at high temperature ($>$1200$^\circ$C), for times on the order of hours, sometimes under applied electric fields \cite{VanOostrom:AoSS:1979,Yamamoto:V:1990,Vancil:ITED:2014,Vancil:2IIVECI:2016}.

B- and M-type cathodes have been reliably and controllably manufactured for decades, yielding cathodes with predictable, repeatable performance in VED applications.
In contrast, reliable manufacture of scandate cathodes with predictable and repeatable performance remains elusive, despite extensive study.
This has limited the adoption of scandate cathodes in VEDs, and motivated extensive and ongoing efforts to develop optimized and repeatable approaches for fabricating high-performing scandate cathodes.
In turn these fabrication efforts have led to diverse reports of the properties and performance of scandate cathodes.

\section{Fabrication, Microstructure, and Emission Performance of Scandate Cathodes}
\subsection{Fabrication}
Multiple research groups over a period of decades have worked to develop fabrication processes that reliably yield consistent high-performing scandate cathodes.
The methods employed by these groups can be broadly categorized by the stage of processing during which Sc is incorporated into the cathode.

\subsubsection{Sc-in-W}
In this approach, sometimes referred to as the ``doped powder" or ``scandia-doped impregnated (SDI)" method in the literature, Sc$_2$O$_3$-doped W powder is produced and then die-pressed into a near-net shape pellet. 
Precursor powders (and, therefore, the subsequent pressed pellets) generally contain a mix of on the order of 1 $\upmu$m W particles decorated with nanoscale particulates of Sc$_2$O$_3$.
A few prominent techniques have emerged to produce Sc$_2$O$_3$-doped W precursor powders.

Two precipitation-based techniques have been explored for producing Sc$_2$O$_3$-doped W powders: ``liquid-solid'' (L-S) \cite{Wang:ASS:2003,Tao:VE:2003} and ``liquid-liquid'' (L-L) \cite{Yuan:ASS:2005,Wang:ITED:2007,Wang:ITED:2009,Wang:IJN:2013,Zhao:ITED:2011,Zhao:TST:2011} approaches.
Both methods yield more evenly dispersed distributions of Sc$_2$O$_3$ particulates compared to mechanical mixing of separately-fabricated W and Sc$_2$O$_3$ powders \cite{Wang:ITED:2007}.
More uniform distributions of Sc-containing material have been reported to improve cathode performance, but the exact mechanism and extent of the effect is unclear, with some studies suggesting that microstructure may be the critical feature \cite{Wang:ITED:2007,Yuan:ASS:2005,Wang:ITED:2009,Yang:ITED:2016}.

In the L-S method, also referred to as the single precipitation approach, Sc$_2$O$_3$ is precipitated from an aqueous solution of ScN (``liquid'') directly onto W or WO$_3$ powder (``solid'') added to the solution.
This yields Sc$_2$O$_3$-doped WO$_x$ that is reduced in a hydrogen atmosphere to form Sc-doped W powder \cite{Wang:ASS:2003,Tao:VE:2003,Yang:ITED:2016}.
The L-L method is a co-precipitation process wherein both Sc$_2$O$_3$ and W are precipitated from an aqueous solution, typically in the form of ScN and ammonium metatungstate, respectively.
The solution is partially dried to form a gel, air-fired to remove organic impurities, then reduced in a hydrogen atmosphere, again yielding Sc-doped or Sc$_2$O$_3$-covered W particles \cite{Wang:ITED:2007,Yuan:ASS:2005,Wang:ITED:2009,Yang:ITED:2016,Wang:AM:2016}.

Sc-containing W pellets formed from L-L or L-S synthesized powders are sintered to achieve mechanical stability and a target final density well below 100\% (20-24\% as reported in Ref.~\cite{Yuan:ASS:2005}, $\sim$60\% as reported in Refs.~\cite{Vancil:ITED:2014,Vancil:ITED:2018}.
Sintered pellets are impregnated by a carbonate mixture \cite{Jiang:I25IVESCPICN:2004}, a  BaO-CaO-Al$_2$O$_3$ mixture \cite{Zhang:FML:2013}, or Ba-Ca aluminates \cite{Vancil:ITED:2014,Liu:ITED:2011,Wang:ITED:2015} calcining, melting, and infiltrating the pellet pores.
Typical Ba, Ca, Al mole ratios in the impregnant powder mix are 4:1:1, 6:1:2, 5:3:2, respectively, similar to those used for B- and M-type cathodes \cite{Yamamoto:RPP:2006,Wang:ITED:2015,Wang:AM:2016,Koppius::,Hasker:ASS:1986}.
Post-impregnation, residual impregnant covering the pellet surface is removed mechanically, and the cathode is washed with deionized water \cite{Wang:IJN:2013,Wang:AM:2016}.
Activation of L-L or L-S cathodes typically consists of heating above $\sim$1150$^\circ$C$_b$ for a period on the order of hours\cite{Wang:IJN:2013,Wang:AM:2016}.
A detailed summary of Wang et al.'s work on both L-L and L-S powder processing and cathode manufacturing can be found in Ref.~\cite{Gaertner:MDiVES:2020}.

\subsubsection{Sc-in-impregnant}
This approach, first implemented by van Stratum et al. \cite{VanStratum::} and Koppius et al. \cite{Koppius::} involves adding Sc$_2$O$_3$ to the impregnant powder mixture, rather than as a dopant or additive in the W pellet. 
Both of these groups investigated several impregnant mixtures containing some combination of Ba, Ca, Al, Sc, and Y.
In these mixtures, Sc$_2$O$_3$ represented approximately 3 wt.\% of the impregnant. 
Impregnation proceeds as for B-, M- and Sc-in-W cathodes, and Sc-in-impregnant cathodes are activated similarly to other scandate cathodes.
Yin et al. developed an impregnant composition of Ba, Ca, Al, Sc, and Sr oxides and reported current densities about 5$\times$ higher than M-type cathodes, although with less emission uniformity \cite{Yin:ITED:2013}.

\subsubsection{Sc-in-oxides}
A related approach implemented by Cui et al.\cite{Cui:ASS:2011,Cui:JRE:2010,Cui:RCI:2011} prepares Sc$_2$O$_3$-doped pressed cathodes by fabricating Sc$_2$O$_3$ and Ba-Ca-aluminate co-doped W powders using a spray drying method and a two-step hydrogen reduction process.
In this ``Ba scandate'' or ``scandia-doped powder (SDP)" fabrication approach, cathode pellets are pressed and sintered from a mixture of W powder and various Ba$_x$Sc$_y$O$_{x+\frac{3}{2}y}$ powders \cite{Wang:ITED:2009,Wang:JoRE:2010}.
This approach does away with a traditional impregnation step by directly incorporating the impregnant precursor oxides in the pressed pellet.
Typical pellets are pressed from mixtures of $\sim$10 wt.\% Ba scandate powder and $\sim$90 wt.\% fine grained W powder. 
Pellets were sintered at $\sim$1570$^\circ$C in a hydrogen atmosphere for 15 minutes then polished with Al$_2$O$_3$ powder and ultrasonically cleaned in Freon.

\subsubsection{Top layer}
Introduced by Hasker, van Esdonk, and Crombeen in 1986 \cite{Hasker:ASS:1986}, ``top layer" cathodes originally consisted of a pure W pellet sputter-coated with W+Sc$_2$O$_3$ or W+W/ScH$_2$. 
Subsequent efforts were extended by impregnating W/Sc-coated porous W pellets in the same manner as other B-, M-, and scandate cathodes---that is, by melting BaCO$_3$, CaCO$_3$, and Al$_2$O$_3$ into the (coated) pellet pores \cite{Gibson:ITED:1989,Gartner:ASS:1997}.
These later approaches also employed laser ablation deposition (LAD) in place of sputter-coating \cite{Gartner:ASS:1997}.
Among examples of these fabrication approaches, the thickness of the Sc-containing ``top layer'' has been widely varied, with studies of cathodes with coating thicknesses ranging from angstrom- \cite{Gibson:ITED:1989} to micron-scale \cite{Wang:ASS:1999} reported. 
The applied Sc-containing coatings have also varied with respect to chemical composition. 
Initially, W+ScH$_2$, W+Sc$_2$O$_3$, or W+Sc$_2$W$_3$O$_{12}$ were used \cite{Gartner:ASS:1997,Hasker:ASS:1986,Yamamoto:JJAP:1988,Yamamoto:ASS:1984}. 
Additional compositions including mixtures with Re, e.g. W+Re$_2$Sc and Re$_{24}$Sc$_5$, have also been explored \cite{Gartner:ASS:1997,Hasker:ITED:1990}.
A thorough survey of G{\" a}rtner et al.'s developments in top layer scandate cathodes can be found in Ref.~\cite{Gaertner:MDiVES:2020}.

\subsection{Microstructure}
Despite significant recent efforts to identify fabrication methods that reliably yield high-performing Sc-containing cathodes, few comprehensive reports of the microstructure of fabricated cathodes are available in the literature.
Interpretation of available results is hampered by the challenges of reliably determining the composition and phase of heterogeneous sub-micron surface features and the compositional complexity of cathode surfaces.
As all major Sc-containing cathode fabrication methods involve loosely pressed and sintered pellets of W powders, key microstructural features of interest include W particle shapes and sizes, the distribution of Sc, and the presence and distribution of impregnant-related species.

\begin{figure}
	\begin{center}
		\includegraphics[width=1\linewidth]{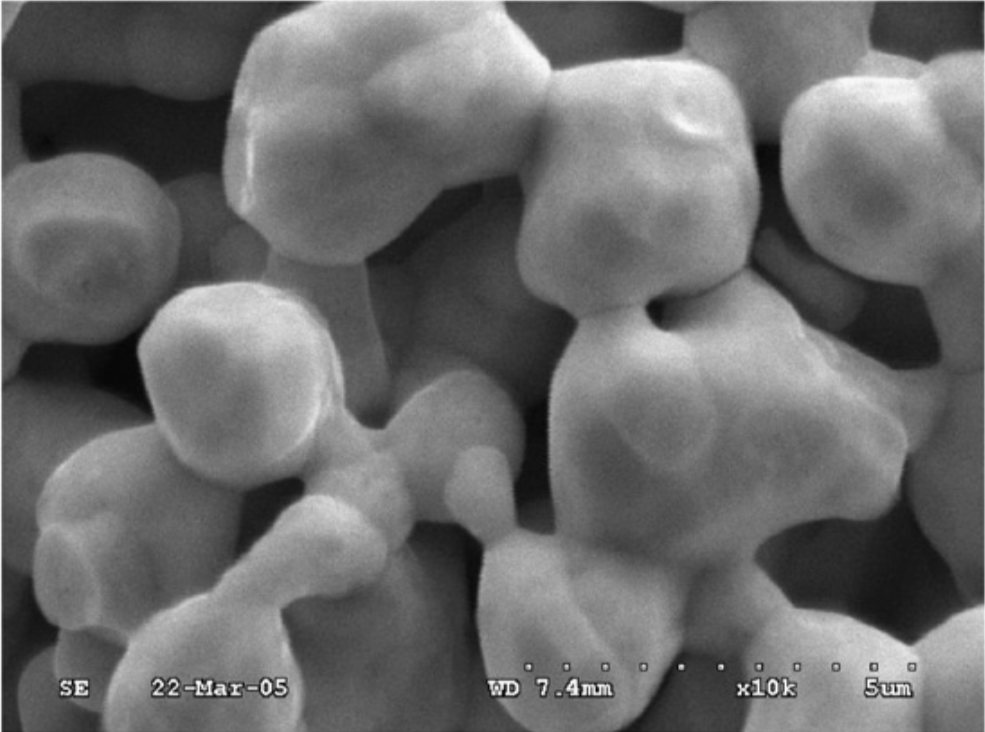}
	\end{center}
	\caption{An SEM image of the impregnated cathode surface; note that it is quite difficult to discern distinct Sc$_2$O$_3$ particles \cite{Wang:JPCS:2008}.}
	\label{fig:impregnated-cathode}
\end{figure}

Wang et al. \cite{Wang:JPCS:2008} prepared Sc-in-W cathodes using L-L powders and reported a resulting microstructure of sub-micron quasi-spherical W grains (see Fig.~\ref{fig:impregnated-cathode}). 
The authors posited that a uniform ``active'' (electron emitting) layer atop the underlying pellet microstructure resulted in its excellent observed emitting properties and that small, near spherical W grains in the pellet (with diameters on the order of 1 $\upmu$m or less) contributed to the formation of this layer.

\begin{figure}
	\begin{center}
		\includegraphics[width=1\linewidth]{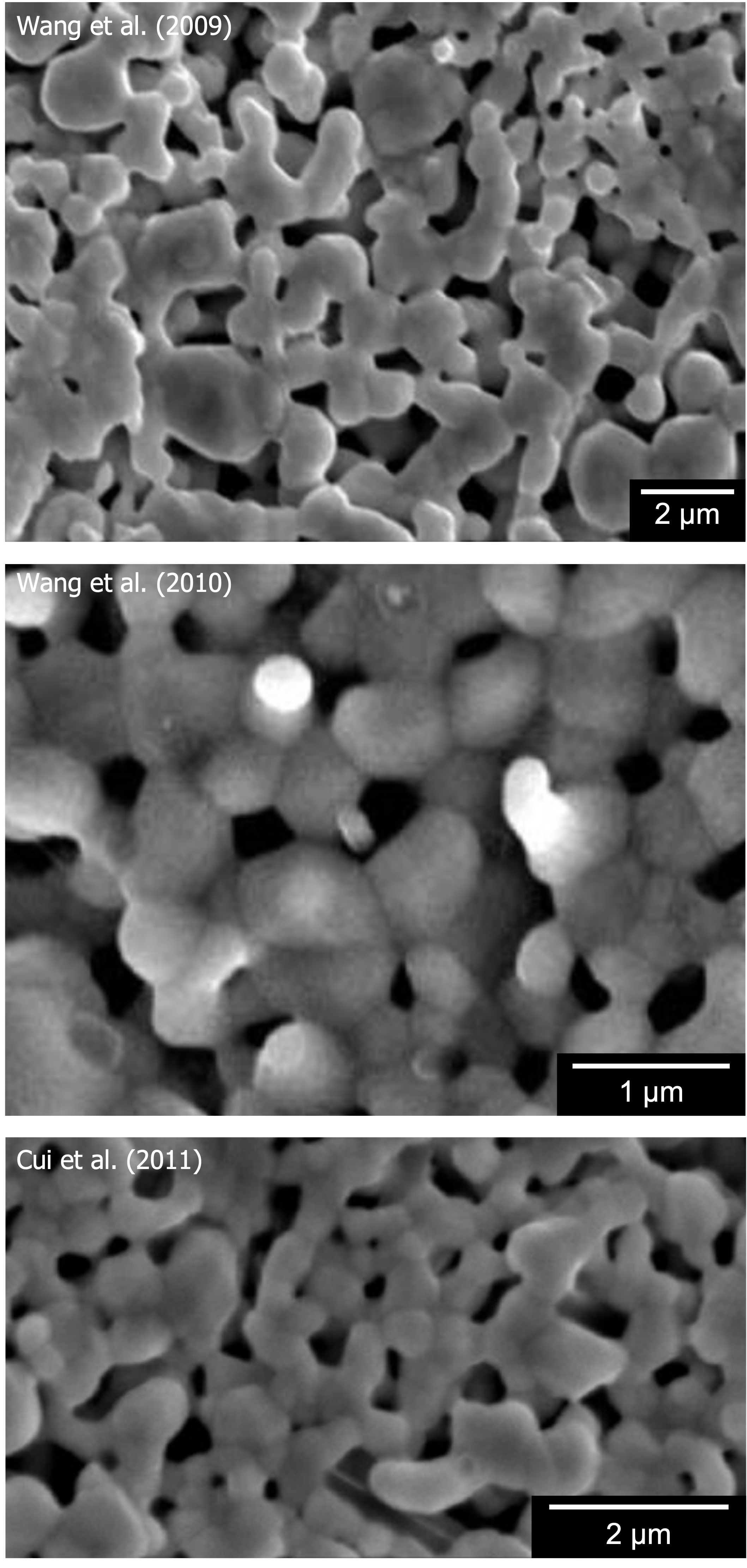}
	\end{center}
	\caption{SEM images of Sc$_2$O$_3$-doped pressed cathodes prepared for different investigations: top (Wang et al., 2009 \cite{Wang:ITED:2009}), middle (Wang et al., 2010 \cite{Wang:JoRE:2010}), bottom (Cui et al., 2011 \cite{Cui:ASS:2011}).}
	\label{scandia-pressed-cathodes}
\end{figure}

Wang et al.\cite{Wang:ITED:2009,Wang:JoRE:2010} also prepared Sc-in-oxides cathodes by die-pressing and sintering Sc$_2$O$_3$-doped W powders mixed with Ba-Ca-aluminates in a hydrogen atmosphere.
Utilizing scanning electron microscopy (SEM), Wang et al. reported a homogeneous resulting porous structure [Fig.~\ref{scandia-pressed-cathodes}(top) and (middle)], with W grains smaller than 1 $\upmu$m. 
The authors posited that this structure is favorable for the active substances (that is, substances that enhanced electron emission) to diffuse from pores in the cathode interior to its surface. 
Some grains with rectangular shape were observed on the surface of the W grains and were reported to be composed of Sc$_2$W$_3$O$_{12}$. 
Cui et al.\cite{Cui:ASS:2011} used similar fabrication methods, and obtained the microstructure shown in Fig.~\ref{scandia-pressed-cathodes} (bottom). 
The surface was reported to be comprised of sub-micron W grains with a homogeneous distribution of Sc$_2$O$_3$ and Ba-Ca-aluminate.
No discrete particles of either Sc$_2$O$_3$ or impregnant oxides were observed.
Compositional analysis via EDS showed that Sc$_2$O$_3$ and Ba-Ca-aluminates were present at the cathode surface. 
Sc was observed at all sampled locations, in some spots with accompanying signals from W, Ba, Ca, and Al, and at others with only accompanying W signals. 
As proposed by Wang et al. \cite{Wang:ITED:2009,Wang:JoRE:2010}, Cui et al. suggested that uniform, equiaxed sub-micron W grains coated in Sc/Ba/O represent an optimal microstructure for high current density thermionic emission.

\begin{figure}
    \centering
    \includegraphics[width=1\linewidth]{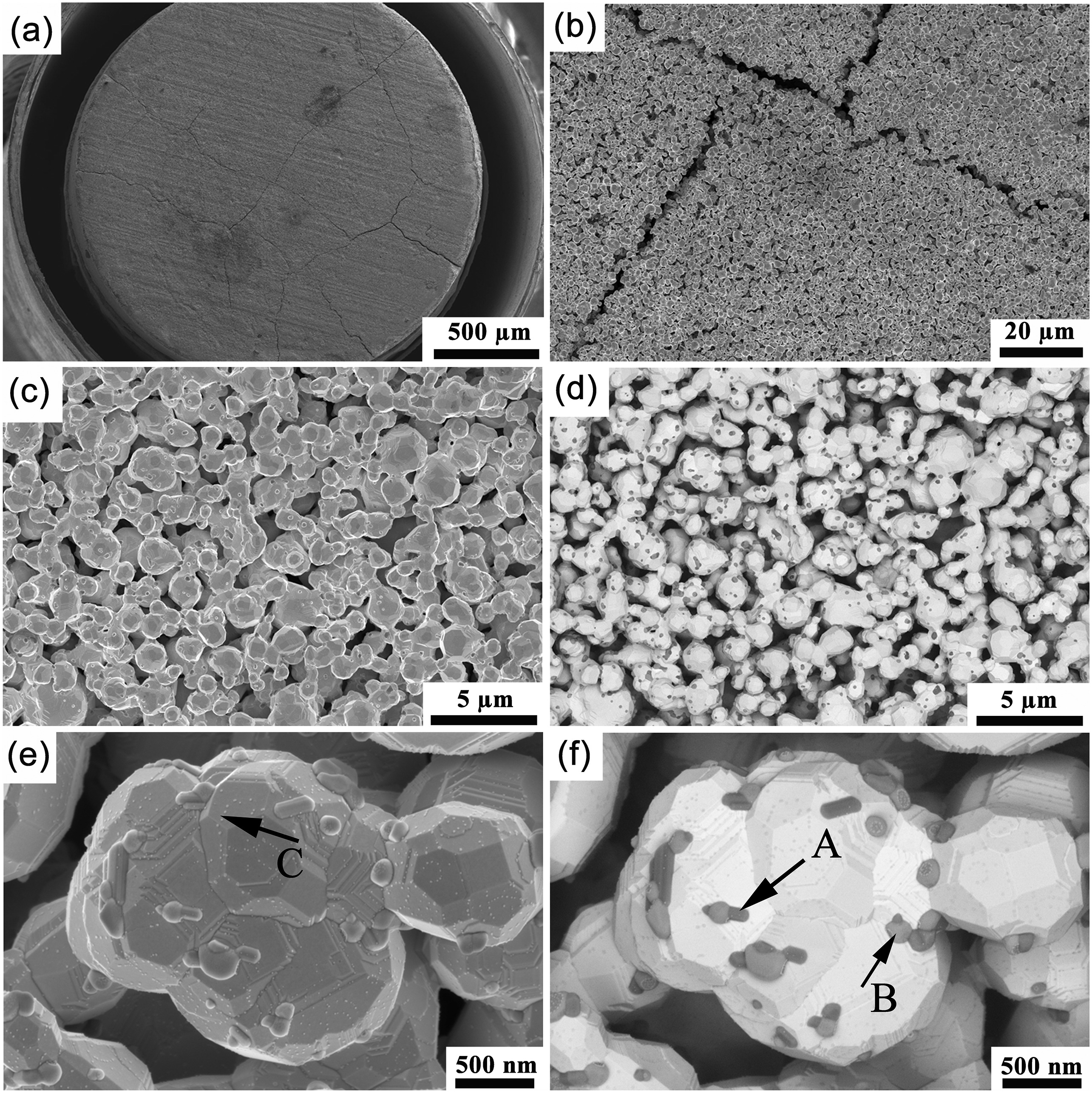}
    \caption{Impregnated scandate cathodes fabricated by Vancil et al. \cite{Vancil:ITED:2014} and characterized by Liu et al. \cite{Liu:MC:2019}. The nanoparticles labeled A and B in (f) are on the order of 100 nm and contrast with the underlying W. These smaller nanoparticles, denoted as C, are on the order of 10 nm and were initially presumed to be Ba/BaO, based on the results reported in Refs.~\cite{Lai:ASS:2018} and \cite{Lamartine:ASS:1986}.}
    \label{fig:balkcathodes}
\end{figure}

Liu et al. \cite{Liu:M:2019,Liu:MC:2019} conducted a thorough study of high-performing Sc-in-W cathodes fabricated with L-S powders.
They reported a microstructure composed of distinct W grains ``decorated" (as opposed to ``coated'') with impregnant oxide particles.
The series of secondary electron and back-scatter electron SEM micrographs shown in Fig.~\ref{fig:balkcathodes} highlight the complex topology of the cathode microstructure as well as the range of phases present in particles on W grains.

\begin{figure}
	\begin{center}
		\includegraphics[width=1\linewidth]{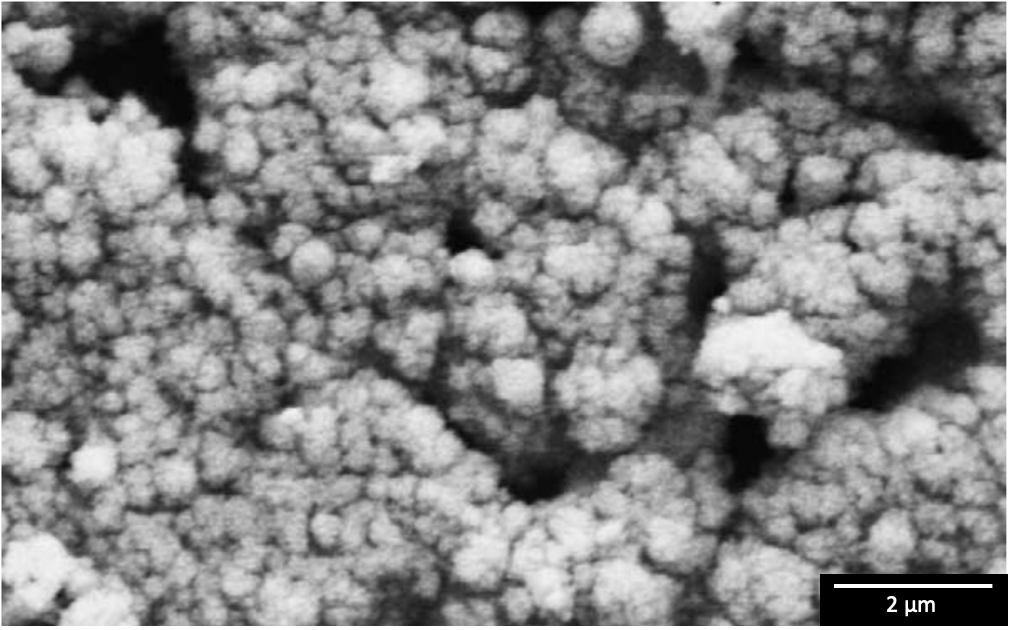}
	\end{center}
	\caption{SEM image of the surface of a LAD top-layer scandate cathode prepared by G{\" a}rtner et al. \cite{Gaertner:MDiVES:2020,Gartner:ASS:1997}.}
	\label{top-layer-cathodes}
\end{figure}

G{\" a}rtner et al. \cite{Gaertner:MDiVES:2020,Gartner:ASS:1997} prepared top-layer scandate cathodes using LAD to deposit a nanocrystalline layer of about 100-500 nm thickness on a porous W matrix of a 4BaO$\cdot$CaO$\cdot$Al$_2$O$_3$ impregnated cathode (``B-type"). 
The resulting microstructure, shown in Fig.~\ref{top-layer-cathodes}, consists of an ultrafine homogeneous distribution of nanoparticles containing Sc$_2$O$_3$ and Re at the surface. Further details of the microstructure and fabrication process are proprietary knowledge of Philips Research Aachen \cite{Gaertner::1993,Gaertner::1993a,Gaertner::1997,Gaertner::2001}. 

\subsection{Emission Performance}
Comparing performance among scandate cathodes fabricated by different researchers is difficult due to the inherent complexity of emission measurement.
Although the emission properties summarized here from reports by a number of different groups are all measured via close-spaced diode (CSD) testing, the details of each individual test configuration are crucial in interpreting and comparing results.
While an example CSD configuration is described by Vancil et al. \cite{Vancil:ITED:2018}, many test set-up and environment details are omitted in all papers referenced here.

Some factors limiting comparability of emitted current measurements include varying anode-cathode distances and applied biases (effectively, the magnitude of the collection field at the cathode surface), the presence or absence of guard rings to limit current collection from the sides of a cathode (as well as other test geometry factors specific to each test set-up), whether the cathode was operated in continuous or pulsed mode, and challenges relating to accurate measurement of the temperature of the cathode.
Temperature measurement is a complicated procedure in the context of hot cathodes. 
Temperatures reported in this section are generally all \emph{brightness} temperatures ($^\circ$C$_b$), not true temperatures. 
Converting from brightness to true temperature requires prior detailed knowledge of the emissivity of the sample surface.
As noted, the composition (and microstructure) of scandate cathodes is complex, not definitively characterized, and potentially dynamic.
Many studies assume that cathode surfaces have the emissivity of pure W.
Although this is generally done out of necessity, it implies that the true temperature of the cathode is not accurately known.

Wang et al. \cite{Wang:ITED:2007} reported that space charge-limited current densities of over 30 A/cm$^2$ at 850$^\circ$C$_b$ were routinely achieved for Sc-in-W, or scandia-doped impregnated, cathodes fabricated from L-S precursor powders. 
This work also reported a current density of 2 A/cm$^2$ at 950$^\circ$C$_b$ over an extremely long lifetime (over 10,000 hours). 
Wang et al. \cite{Wang:JPCS:2008} also report that a Sc-in-W cathode developed from L-L precursor powders reached 52 A/cm$^2$ at 850$^\circ$C$_b$. 
In a study of Sc-in-W cathodes containing Re fabricated with L-S precursor powders, Wang et al. \cite{Wang:JAC:2008} reported high current densities ($>$30 A/cm$^2$) at 850$^\circ$C$_b$, comparable to the group's measurements for Sc-in-W cathodes without Re.
Wang and Liu et al. \cite{Liu:ASS:2005,Wang:ITED:2009,Wang:ITED:2007,Wang:JAC:2008} attributed the enhanced emission capabilities of these cathodes to a $\sim$100 nm Ba-Sc-O multilayer that coats the W grains.

Vancil et al. \cite{Vancil:ITED:2018} investigated the correlation between ``knee temperature''---defined as the temperature at which electron emission shifts from space-charge limited to temperature limited---with aspects of precursor powder fabrication. 
Knee temperature is a proxy for emission performance, as lower knee temperatures imply higher thermionic current densities at lower temperatures.
Cathodes composed of L-L powders formed during a later-stage of precipitation (referred to as ``S" fraction in the reference) and later hydrogen-fired were reported to exhibit the lowest knee temperatures---between 802$^\circ$C$_b$ and 845$^\circ$C$_b$. 
Minor variation in powder preparation, such as slight changes in the precipitation or drying methods, were reported to shift knee temperatures significantly--up to as high as 944$^\circ$C$_b$, implying lower emitted current densities.

For Sc-in-impregnant cathodes (impregnated with a combination of BaO, Sc$_2$O$_3$, CaO, and Y$_2$O$_3$), Koppius \cite{Koppius::} reported a current density of up to 5 A/cm$^2$ at 1000$^\circ$C.
These authors also reported that the evaporation of metallic Ba from the cathode surface (an issue hypothesized to reduce the lifetime of B- and M-type cathodes) is very slow, and that cathode lifetime approached 2,000 hours. 
Van Stratum \cite{VanStratum::} evaluated similarly composed and fabricated cathodes and reported comparable emission properties. 
Sc-in-impregnant cathodes with different impregnant compositions---BaO/Sc$_2$O$_3$ and BaO/CaO/Al$_2$O$_3$/Sc$_2$O$_3$---were also considered.
For the BaO+Sc$_2$O$_3$ impregnant, Van Stratum et al. \cite{VanStratum::} reported cathodes exhibiting a current density of 1.5-4 A/cm$^2$ at 1100$^\circ$C and a lifetime of 2,000--3,000 hours. 
The BaO/CaO/Al$_2$O$_3$/Sc$_2$O$_3$-impregnated cathode was reported to have a current density of 5 A/cm$^2$ at 1000$^\circ$C and a lifetime of 3,000 hours. 
It should be noted that the temperature measurement approach of both Ref.~\cite{Koppius::} and \cite{VanStratum::} was not reported, though another paper published by van Stratum and Kuin \cite{VanStratum:JAP:1971} that focused on M-type cathodes measured brightness temperature with an optical pyrometer. 

Van Oostrom and Augustus \cite{VanOostrom:AoSS:1979} reported that Sc-in-oxides cathodes (referred to in the reference as ``Ba scandate'' cathodes), fabricated with a mixture of 3BaO$\cdot$2Sc$_2$O$_3$ as discussed in Ref.~\cite{Komissarova:ZNKUETSRJICET:1965}, exhibited a current density of $\sim$10 A/cm$^2$ at 950$^\circ$C$_b$.
Cui et al. \cite{Cui:ASS:2011} also reported high emission---a current density of 31.5 A/cm$^2$ at 850$^\circ$C$_b$---for similarly fabricated cathodes. 
Wang et al. \cite{Wang:ITED:2009} characterized the emission properties of Sc-in-oxides cathodes prepared by different methods -- mechanical mixing, high-energy ball milling, and L-L doping. 
This study reported that cathodes prepared by the L-L method exhibit superior emission properties---a current density of 46 A/cm$^2$ at 850$^\circ$C$_b$. 
Van Oostrom and Augustus \cite{VanOostrom:AoSS:1979} attributed the enhanced emission properties of Sc-in-oxides cathodes to activated Sc$_2$O$_3$ regions on the cathode surface, while Cui et al. \cite{Cui:ASS:2011} credited it to an active Ba-Sc-O multilayer. 

In addition to Sc-in-W and Sc-in-oxides cathodes, ``top layer'' scandate cathodes have also been reported to exhibit low work functions: 1.45-1.5 eV at $\sim$1100 K (830$^\circ$C) \cite{Gibson:ITED:1989}. 
Top layer scandate cathodes have been reported to have long lifetimes relative to Sc-in-oxides cathodes and better recovery capability after ion bombardment than Sc-in-W cathodes \cite{Hasker:ASS:1986}.
G{\" a}rtner et al. \cite{Gartner:ASS:1997} reported that a top layer W/Re+Sc$_2$O$_3$ scandate cathode prepared by LAD exhibited a high current density---zero field emission of 400 A/cm$^2$ at 965$^\circ$C$_b$ and 32 A/cm$^2$ at 760$^\circ$C$_b$.
Hasker et al. \cite{Hasker:ASS:1986,Hasker:ITED:1990} posited that a top layer comprised of partially Sc$_2$O$_3$-covered W grains improved emission homogeneity and ion bombardment resistance. 
Furthermore, it was claimed that, for top layer scandate cathodes, emission mainly stems from BaO globules on the W grains rather than metallic Ba or Ba-O on bulk Sc$_2$O$_3$ \cite{Hasker:ASS:1986}.
Uda et al. \cite{Uda:ASS:1999} studied the thermionic emission of top-layer scandate cathodes with varied coating thickness.
This work reported that thinner top layers of Sc$_2$O$_3$ (2 nm) and W (8 nm) exhibited pulse emission of $\sim$80 A/cm$^2$ at 1300 K ($\sim$1030$^\circ$C), an improvement over thicker coatings. 

\section{Working Models for Mechanisms of Enhancing Emission} \label{roleofsc}
\subsection{Introduction}
Despite diverse fabrication approaches, complex microstructures that are challenging to interpret, and limitations in directly comparing reports of emission performance, it is widely accepted that scandate cathodes have demonstrated the potential to yield up to 10-fold increases in thermionic current density at operating temperatures reduced by $\sim$100$^{\circ}$C compared to industry-standard M-type cathodes.
As reliable fabrication approaches have remained elusive, extensive efforts to identify the specific mechanism(s) by which the addition of Sc enhances emission performance have been undertaken.
Several working models have been proposed for mechanisms by which Sc influences thermionic emission.
In the \emph{Ba adsorbed on Sc-containing layer} model a Sc-containing layer atop W is proposed to enhance emission in the same way as Os-Ru layers in M-type cathodes \cite{Lesny:ITED:1990}. 
In the \emph{semiconductor} model a Sc-containing semiconducting layer is proposed to form, allowing the anode-cathode field to penetrate the surface of the cathode, reducing the apparent work function \cite{Raju:ITED:1994,Maloney:ASS:1985}.
Finally, in the \emph{dipole} model Sc is understood to play a role in enhancing or stabilizing either an intrinsically polar surface structure or a layer of adsorbed polar complexes.
Surface dipoles alter the potential barrier for electron emission, reducing the apparent work function\cite{Cortenraad:ASS:2002}.
All of these models have been adapted from models previously applied to B-type, M-type, or oxide cathodes.


\subsection{Ba adsorbed on Sc-containing layer model}\label{adsorptionmodel}
The Ba adsorbed on Sc-containing layer model posits that a multilayer of Sc$_2$O$_3$-containing material forms on W surfaces.
Ba or BaO then adsorbs on the Sc-containing layer, yielding low work functions. 
This is directly analogous to M-type cathodes, where Ba or BaO adsorbs atop a metallic Os-Ru coating on W grains. 
In this mechanism, the role of W is merely to serve as a conducting path to deliver
electrons to be emitted from surfaces consisting of BaO atop a Sc$_2$O$_3$-containing layer.

\begin{figure}
	\centering
	\includegraphics[width=0.5\textwidth]{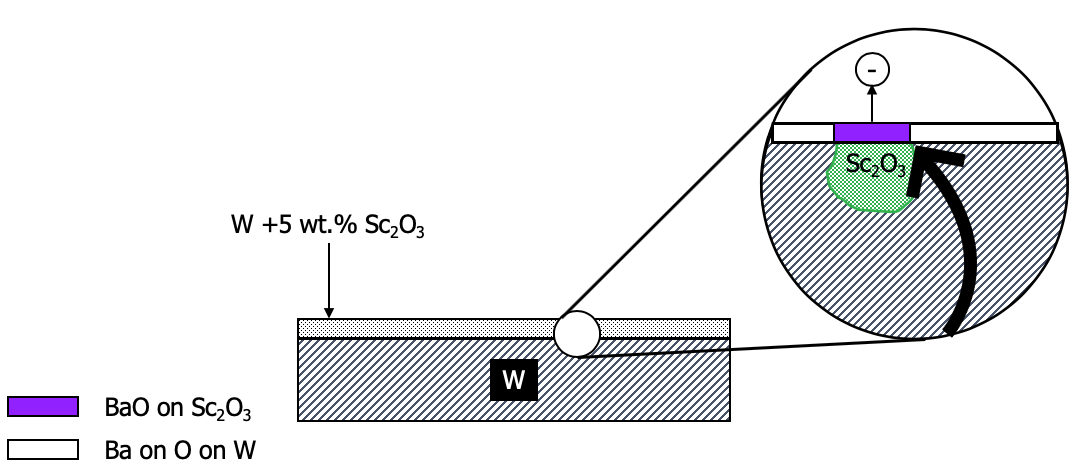}
	\caption{The working model for the "top-layer" scandate cathodes, as suggested by Lesny and Forman \cite{Lesny:ITED:1990}. On the left, the structure of the cathode is illustrated; on the right, an exploded view showing the thin layer surface configuration. This is a reproduction of the figure originally included in the reference.}
	\label{fig:lesnyformanmodel}
\end{figure}

This model was first proposed for scandate cathodes by van Oostrom and Augustus \cite{VanOostrom:AoSS:1979} in 1979 based on Auger spectroscopy studies of Sc-in-oxides cathodes that detected Ba, Sc, O, and W at the cathode surface after activation, with both Ba and Sc exhibiting binding energy shifts characteristic of oxides. 
After sputter-cleaning the cathode surfaces via Ar ion bombardment, only W, Sc, and O were detected, suggesting that BaO was only present in a thin layer atop Sc$_2$O$_3$ on W.
A decade later, Lesny and Forman \cite{Lesny:ITED:1990}, focusing on cathodes fabricated via the ``top layer'' approach, reached similar conclusions. 
They studied high-temperature emission and desorption of Ba from a W-foil-based model system fabricated to simulate the chemical conditions at scandate cathode surfaces.
Samples on W foils were fabricated by first depositing metallic Sc then introducing O$_2$ to oxidize the Sc.
Ba was then deposited on the outermost surface, followed again by O$_2$ to oxidize the Ba. 
Samples were heated in a diode test configuration and the thermionic emission measured.
Emission current was observed to be maximized when Auger detected both BaO and Sc$_2$O$_3$ at the surface, as opposed to either Ba and Sc$_2$O$_3$ or just BaO.
Lesney and Forman concluded that the optimal surface structure for scandate cathodes consisted of on the order of a monolayer of BaO adsorbed on Sc$_2$O$_3$ (as shown in Fig.~\ref{fig:lesnyformanmodel}). 

In an effort to identify the optimal composition of Sc-containing layers proposed to form atop W, Magnus \cite{Magnus:ASS:1997} examined the thermionic emission behavior at 925-1125$^\circ$C for nine compounds from the BaO-Sc$_2$O$_3$-WO$_3$ ternary system deposited on W: Sc$_6$WO$_{12}$, Sc$_2$W$_3$O$_{12}$, BaWO$_4$, Ba$_2$WO$_5$, Ba$_3$WO$_6$, BaSc$_2$O$_4$, Ba$_3$Sc$_4$O$_9$, Ba$_2$Sc$_2$O$_5$, and Ba$_3$Sc$_2$WO$_9$. 
In all cases, coating these oxide materials with BaO enhanced emission, and measured emission currents indicated that a mixture of Ba$_3$Sc$_2$WO$_9$ with 10 wt.\% W powder and, separately, the compounds Sc$_2$W$_3$O$_{12}$ and Sc$_6$WO$_{12}$ all behaved consistent with previous reports of scandate cathodes. 
Of these compounds, mixtures of Ba$_3$Sc$_2$WO$_9$ with Sc$_2$O$_3$ were found to exhibit the most promising emission characteristics, and the enhancing effects of added BaO were taken to imply that optimal surfaces consisted of BaO adsorbed atop the oxide \cite{Magnus:ASS:1997, Magnus:ASS:1997a}.

Several studies by Vaughn and Kordesch \cite{Vaughn:ITED:2009,Kordesch:JVSTB:2011,Vaughn:2IIVECI:2010,Vaughn::2010} also support the Sc-containing layer model.
Like Lesny and Forman \cite{Lesny:ITED:1990}, these authors studied emission properties of various model surfaces synthesized to reproduce cathode systems---BaO on W, Sc$_2$O$_3$ on W, Sc and Ba co-sputtered on W, Sc$_2$O$_3$ on BaO on W, and BaO on Sc$_2$O$_3$ on W. 
Thermionic electron emission microscopy (ThEEM) and photoelectron emission microscopy (PEEM), described in Ref.~\cite{Wan:U:2012}, were used to image the electron emitting regions and correlate brightly emitting regions with areas of W films coated with the dissimilar materials.
While the emission behavior and deposited layers of many samples were observed to evolve with time at high temperature, Vaughn and Kordesch reported that the BaO/Sc$_2$O$_3$/W system was both stable and the only multilayer system that showed thermionic emission comparable to that reported for scandate cathodes.
These findings further support a model for enhanced emission from scandate cathodes as resulting from BaO adsorbed on an Sc$_2$O$_3$ layer coating W grains.

Using density functional theory (DFT), Jacobs et al. \cite{Jacobs:JPCC:2014} studied the thermodynamic stability and work function of Ba/O dimers adsorbed on the (011) and (111) crystallographic surfaces of crystalline Sc$_2$O$_3$. 
This computational model system was designed to mimic the surface features active in the Sc-layer model and computed results showed that Ba-O dimers on (111) surfaces of crystalline Sc$_2$O$_3$ were \emph{not} stable over the full range of surface coverages up to one monolayer relative to the formation of a (bulk) crystalline BaO layer.
In contrast, a stable configuration of 7 Ba-O dimers adsorbed per unit surface of (110) Sc$_2$O$_3$ ($\sim$0.583 dimers per surface O) was reported to yield a work function of 1.21 eV, comparable to effective work functions reported for scandate cathodes. 
The Ba-O dimer was suggested to be both a surface dipole that induced surface charge rearrangement, thereby modifying the electrostatic potential at the surface and a doping component that changed the Fermi level of the semiconductor Sc$_2$O$_3$ and subsequently modified the work function \cite{Jacobs:JPCC:2014}.


Although these studies were generally used to support the idea of Ba-containing species adsorbed atop an identifiable layer of Sc or Sc-containing material, they do not directly probe cathode surface configurations. 
Therefore, whether Sc (or Sc$_2$O$_3$) remains as an identifiable layer with Ba (or BaO) adsorbed atop it after activation and/or operation cannot be determined, even when Ba and Sc are deposited separately and sequentially. 
Computationally, Jacobs et al. \cite{Jacobs:JPCC:2014} reported that Ba-O/Sc$_2$O$_3$ is both stable and exhibits a low work function. 
However, this system required a bulk-like Sc$_2$O$_3$ layer with a layer thickness on the order of nm, much greater than a monolayer.
The conductivity of such an oxide layer would then be a concern, and, to date, no thick layers of Sc$_2$O$_3$ on cathode surfaces have been observed \cite{Liu:MC:2019}.

\subsection{Semiconductor Model} \label{semiconductormodel}
\begin{figure}
	\centering
	\includegraphics[width=0.5\textwidth]{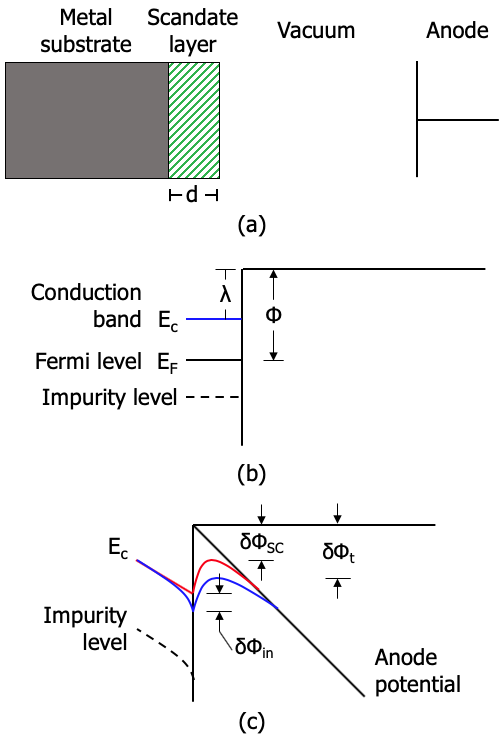}
	\caption{Energy level diagram that depicts the semiconducting behavior of a scandate cathode: (a) scandate cathode in a plane parallel diode, (b) energy levels absent an external field (where $\lambda$ is the electron affinity), and (c) energy levels modified by the anode at the surface. With an applied field, the total work function reduction ($\delta \Phi_t$) is equivalent to $\delta \Phi_t$ = $\delta \Phi_{in}$ + $\delta\Phi_{SC}$. This is an adaption of the figure originally included in  Ref.~\cite{Raju:ITED:1994}.}
	\label{fig:semiconductormodel}
\end{figure}

The semiconductor model is widely applied for oxide cathodes, where, at elevated temperature, the emitting oxide material is treated as a wide band gap semiconductor \cite{Morgulis:JP:1946,Wright:PPSSB:1952}. 
Unlike metals, which screen electric fields, semiconductors allow external electric fields to penetrate through the surface and into the material.
This effect causes an internal work function reduction \cite{Raju:ITED:1994}, a  phenomenon distinct from the external work function reduction induced by the Schottky effect \cite{Simmons:PRL:1965}. 
As a result, the total reduction (due, e.g., to the anode-cathode field) in the work function of a semiconductor is the sum of the internal and external reduction, denoted as $\delta\Phi_{\textrm{in}}$ and $\delta\Phi_{\textrm{SC}}$, respectively (Fig. \ref{fig:semiconductormodel}).
The semiconductor model was adapted by Raju and Maloney \cite{Raju:ITED:1994,Maloney:ASS:1985} and applied to scandate cathodes based on experimental evidence suggesting that scandate cathodes were more susceptible to field enhancement than previous dispenser cathodes. 
Compared to B-type cathodes, electron emission in scandate cathodes increases more significantly as the anode field strength increases. 
This ``anomalous'' phenomenon is a deviation from the Schottky law, and can be explained by the presence of a semiconducting surface layer through which the applied field can penetrate.

Given a fixed external field strength (and therefore a fixed $\delta\Phi_{\textrm{SC}}$) the apparent work function of a semiconducting layer is controlled by $\delta\Phi_{\textrm{in}}$, which is determined by the thickness, electron density, and dielectric constant of the semiconductor material. 
In the original semiconductor models proposed by Morgulis \cite{Morgulis:JP:1946} and Wright and Woods \cite{Wright:PPSSB:1952}, the semiconductor layer was considered to be infinitely thick. 
This assumption is reasonable for oxide cathodes, but failed to match detailed measurements of emission from scandate cathodes \cite{Raju:ITED:1994}.
This discrepancy was attributed to the presence of a finite thickness semiconductor layer in scandate cathodes, and Raju et al. \cite{Raju:ITED:1994} modified the Wright-Woods model to account for this.
For thin semiconductor layers the electric field can only penetrate the emitter surface up to the metal substrate, and the internal reduction in the work functions due to an applied external field is expressed as:
\begin{equation}
	\text{sinh\textbf{e}}\Big(\frac{\delta\Phi_{\textrm{R}}}{\textrm{2k$_{\textrm{B}}$}T}\Big) \Bigm/ \textrm{sinh\textbf{e}}\Big(\frac{\delta\Phi_{\textrm{W}}}{\textrm{2k$_{\textrm{B}}$}T}\Big) = \textrm{tanh}\Big(\frac{d}{L}\Big)	\label{WFreduction}
\end{equation}
Here \textbf{e}, $T$, and k$_\textrm{B}$ are the electron charge, temperature, and Boltzmann’s constant, respectively.
The parameters $\delta\Phi_{\textrm{R}}$ and $\delta\Phi_{\textrm{W}}$ are the internal work function reduction corresponding to the Raju (finite thickness) and Wright-Woods (infinite thickness) models, respectively. 
The variables $d$ and $L$ are the thickness and Debye length of the semiconductor layer. 
As indicated from Eq.~\ref{WFreduction}, when the semiconductor layer becomes very thick, tanh($d$/$L$) $\rightarrow$ 1, and the Raju and Wright-Woods models are equivalent. 
When the semiconductor layer is thin, $\delta\Phi_{\textrm{R}}$ increases with the thickness $d$. 
Therefore, a thick semiconductor layer is desirable for optimum internal work function reduction. 

In applying the semiconductor model to scandate cathodes, explaining the observed enhanced emission requires either that the thickness ($d$) of the semiconductor layer is large, or that the value of $n_0$/$\epsilon_r$ for the semiconductor layer is small, as L = (2$n_0$e$^2$/$\epsilon_r\epsilon_0$$k_B$$T$)$^{1/2}$.
The variables $n_0$ and $\epsilon_r$ are the electron density in and dielectric constant of the semiconductor layer, respectively.
Using an electron density on the order of 10$^{14}$ electrons/cm$^3$ and a dielectric constant of 10, both of which are typical for BaO layers \cite{Wright:PPSSB:1952}, the required thickness of the semiconductor layer in scandate cathodes has been estimated to be 0.4 $\upmu$m at 1066 K, or 0.5 $\upmu$m at 1013 K in order to agree with experimental results \cite{Raju:ITED:1994}.


The semiconductor model is supported by a number of studies, and a variety of compositions/structures for the semiconductor layer have been proposed. 
Due to the reported detection of W, Ba, Sc, and O at the surface of scandate cathodes using Auger electron spectroscopy (AES), scanning electron microscopy (SEM), and energy dispersive X-ray (EDS) spectroscopy \cite{Raju:ITED:1994,Maloney:ASS:1985}, compounds in the ternary system BaO-Sc$_2$O$_3$-WO$_3$ \cite{Magnus:ASS:1997, Schoenbeck::2005}, along with the binary systems, BaO-W$_3$ \cite{Kreidler:JACS:1972}, BaO-Sc$_2$O$_3$ \cite{Kwestroo:MRB:1974,Kwestroo:MRB:1982}, and Sc$_2$O$_3$-W$_3$ \cite{Halum:JACS:1985}, have been studied. 
In the work of Raju and Maloney \cite{Raju:ITED:1994,Maloney:ASS:1985}, the semiconductor layer was assumed to be composed of Sc$_6$WO$_{12}$ from the Sc$_2$O$_3$-WO$_x$ system \cite{Halum:JACS:1985}, and some compounds from the BaO-Sc$_2$O$_3$ system, such as Ba$_3$Sc$_4$O$_9$, BaSc$_2$O$_4$, and Ba$_2$Sc$_2$O$_5$. 

Based on Ba/BaO chemical shifts measured via AES, Shih et al. \cite{Shih:ASS:2005,Shih:I25IVESCPICN:2004} suggested that a thick layer of BaO-containing semiconducting layer formed atop Sc$_2$O$_3$ at cathode surfaces from reactions among BaO-Sc$_2$O$_3$-WO$_3$. 
The importance of this layer was deduced from the fact that cathode emission was observed to be non-uniform, with bright (more emissive) and dark (less emissive) regions proposed to be associated with patches of thick Ba-Sc-O containing semiconducting layers and bare W regions, respectively.

Wang et al.\cite{Wang:ITED:2007,Wang:JVSTBNMMPMP:2011,Wang:ITED:2009,Wang:ITED:2009a}, who reported scandate cathodes with a knee temperature (transition temperature from temperature- to space-charge-limited emission) as low as $\sim$800$^\circ$C and uniform, high emission on the order of 100 A/cm$^2$, also supported the semiconductor model.
These studies utilized XRD, EDX and \emph{in situ} AES to analyze the chemical compositions of the cathode surface and found that Sc, Ba, and O concentrations at the surface all increased during activation. 
It was assumed that these components diffused to the surface from inside the W matrix after reactions between Sc$_2$O$_3$ and the impregnants resulted in free metallic Ba and Sc. 
AES depth profile analysis showed signals for Ba, Sc, and O down to 100 nm in W grains with atomic concentration ratios of Ba:Sc:O around (1.5-2):1:(2-3). 
Optimum emission was found to correspond to a Ba:Sc:O ratio of around 1.6:1:2.25. 
As a result, the semiconductor layer was proposed to be a uniform layer of Ba-Sc-O on W grains about 100 nm in thickness, significantly thinner than in any hypothesis described by Raju and Maloney \cite{Raju:ITED:1994}.

An important consideration for emission enhancement via the semiconductor model is the electrical conductivity of the semiconductor layer itself, which is a key limitation of emitted electron flux for oxide cathodes due to the low free electron density available in the emitting semiconductor layer \cite{Yamamoto:RPP:2006}. 
Brodie \cite{Brodie:ITED:2011} proposed that a monolayer of Ba atop Sc$_2$O$_3$ nanocrystals served as an electron donor to yield an n-type semiconductor with a maximum electron density of about 2.4$\times$10$^{20}$/cm$^3$ when the size of Sc$_2$O$_3$ crystallites were on the order of 100 nm, providing sufficiently good conductivity and mobility to yield theoretical results consistent with experiment. 
Jacobs et al. \cite{Jacobs:PRB:2012} calculated the conductivity of Sc$_2$O$_3$ with intrinsic and impurity defects and found that, though intrinsic defects alone were not sufficient, a very small ($\sim$7.5$\times$10$^{-3}$ ppm) impurity concentration was adequate to raise the conductivity of Sc$_2$O$_3$ sufficient to bolster thermionic emission in scandate cathodes.



\begin{figure}
	\centering
	\includegraphics[width=1\linewidth]{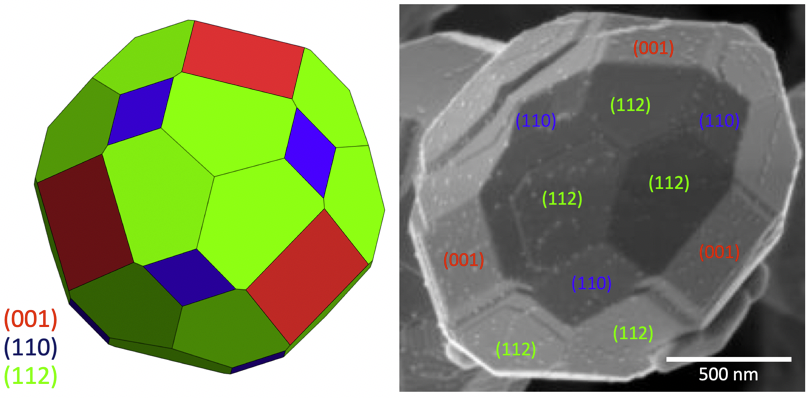}
	\caption{\textbf{(left)} Wulff construction of the equilibrium W grain shape present throughout high-performing scandate cathodes \textbf{(right)} SEM image, reproduced from Ref.~\cite{Liu:MC:2019}, of the characteristic W grain shape. }
	\label{fig:Wwulffshape}
\end{figure}

Using a range of characterization techniques (e.g. SEM, XRD, TEM, and XPS),  recent experimental work on scandate cathodes prepared with both L-S and  L-L powders report results inconsistent with the semiconductor model.
High-resolution SE micrographs showed highly-faceted W grains (on the order of 1 $\upmu$m in length) decorated with nanoparticles (on the order of 10 nm), as well as larger Sc$_2$O$_3$ and Ba-Al-O-containing nanoparticles (on the order of 100 nm) \cite{Liu:MC:2019}. 
Compositional analysis and TEM diffraction performed on cross sections of W grains obtained with FIB both suggested that the W is crystalline with no appreciable oxide/semiconductor layer at the surface. 
The W grains are consistently terminated with (001), (110), and (112) crystallographic planes. 
A semiconductor layer at the surface would likely prohibit such well-defined facets, as shown in Fig.~\ref{fig:Wwulffshape}.

\subsection{Dipole Model}\label{dipolemodel}
\subsubsection{Theory and Background}
\begin{figure}
	\centering
	\includegraphics[width=1\linewidth]{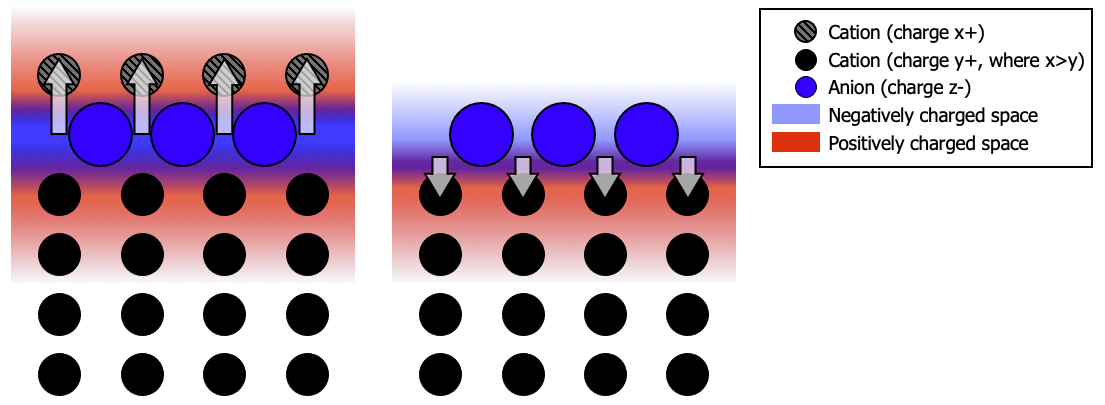}
	\caption{The dipole model posits that the creation of surface dipoles that, when oriented outwards (as in the left construction), lower the electrostatic barrier for electron emission.}
	\label{fig:dipolemodel}
\end{figure}

Perhaps the most widely explored model for lowering work function (and therefore enhancing electron emission) centers on the creation of surface dipoles that, when oriented outwards, lower the electrostatic barrier for electron emission (see Fig.~\ref{fig:dipolemodel})\cite{Nottingham:HdP:1956,Taylor:PR:1933,Ishida:PRB:1988,Ishida:PRB:1990,Soukiassian:PRB:1985,Bonzel:SSR:1988,Fall::1999,Muscat:PiSS:1978}.
In B- and M-type cathodes discrete (and mobile) atomic or molecular species containing Ba cations adsorbed on the emitting surface form the outward facing dipoles \cite{Lesny:ITED:1990,Cortenraad:JoAP:2001,Norman:PRL:1987,Haas:ASS:1985,Ibrahim:PRB:2007,Lai:ASS:2018,Aida:JAP:1993}, and are representative of the broader (and widely studied) phenomenon of WF reduction due to the adsorption of alkali and/or alkali-earth metals\cite{Yamamoto:RPP:2006,Cortenraad:ASS:2002}. 
Sc cations have been proposed to adsorb with Ba cations to give a more optimal density of adsorbed cations and therefore maximize work function reduction.
It has also been proposed that the presence of Sc, while not directly functioning as an adsorbed cation, acts to enhance the binding and/or stability of adsorbed Ba cations, reducing Ba desorption and therefore enhancing work function reduction and/or extending cathode lifetime \cite{Lesny:ITED:1990,Gartner:ASS:1999}.
Separately, various studies have considered the possibility that fixed surface structures containing Sc are formed that are either inherently dipolar or support more optimal densities of adsorbed surface cations.
Finally it has been proposed that the nearby presence of Sc (or Sc-containing compounds) acts on the chemical environment local to emitting surfaces in such a way as to stabilize (or drive formation of) specific surface structures with low work functions.

From a fundamental perspective, the details of how adsorbates form surface dipoles and their quantitative effect on work function have been widely explored.  
In the case of metal-on-metal adsorption there are several models (described in Refs.~\cite{Gurney:PR:1935,Lang:PRB:1971,Lang:PRB:1973,Brodie:SS:2014,Brodie:PRB:1995,Ishida:SS:1991}) proposed to explain the dependence of work function on the adsorption of alkali/alkali-earth metals. 
Gurney \cite{Gurney:PR:1935} suggested that the wave functions of the adatoms and outermost metal substrate atoms overlap, allowing electrons on the adatom to tunnel into the metal, which decreases the electrostatic potential at the surface and reduces work function. 
The dipole strength was dependent on the position of the broadened level relative to the Fermi level. 
Lang \cite{Lang:PRB:1971,Lang:PRB:1978} described the substrate metal using the simple jellium model as a uniform background, and the adsorbed ions as an adjoining uniform positive slab. 
The work function was determined by both the coverage and the thickness of the slab (i.e. ionic separation of adsorbate from substrate).

More recently, evidence has accumulated that the surface dipole forming adsorbates are intrinsically dipolar--that is, that it is Ba-O units adsorbed on B- and M-type cathode surfaces. 
Regardless of the specific nature of the adsorbate-generated dipoles it is widely understood that the work function of a metal substrate decreases with increasing adsorption coverage up to some coverage, beyond which dipole-dipole interactions on the surface limit the effect of additional dipoles, resulting in a plateau in the work function \cite{Taylor:PR:1933,Ishida:PRB:1988,Ishida:PRB:1990,Soukiassian:PRB:1985,Bonzel:SSR:1988}.
Classically, the adsorbed alkali/alkali-earth metals or their oxide complex were considered ions or molecules.
Thus, work function reduction results from the aggregate effect of individual dipoles \cite{Taylor:PR:1933} or induced dipoles caused by surface charge redistribution \cite{Ishida:PRB:1988,Ishida:PRB:1990,Bonzel:SSR:1988,Soukiassian:PRB:1985}. 

\subsubsection{Sc and optimal surface configurations}
The effect of adsorbed (polar) complexes like Ba-O on work function has been reported to be highly dependent on the electronic structure of substrate surfaces---and therefore on the specific crystallographic plane and structure of the substrate surface \cite{Zhou:ASS:2018,Vlahos:PRB:2010,Brodie:PRB:1995,Swartzentruber:JoVS&TA:2014}. 
It has been reported that substrate metals with larger numbers of unoccupied $d$-orbitals bind strongly with Ba-O complexes, enhancing work function reduction for W compared to Pt substrates \cite{Cortenraad:ASS:2002}. 
Similar effects are expected to be at play for the effects of adsorbates on different crystallographic planes of the same substrate \cite{Becker:ANYAS:1954,Webster::1963,Wan:U:2012}.
Brodie\cite{Brodie:PRB:1995} suggested that disparities in work function for different facets is related to direction of the effective mass of an electron with the Fermi energy inside the crystal.

Leveraging significant advancements in quantum mechanical calculation, a large number of computational studies have attempted to identify the specific surface facets and specific surface atomic configurations that optimize the effect of dipoles on work function (and therefore thermionic emission).
In terms of scandate cathodes such studies have explored the hypothesis that adsorbed mixed layers of Ba and/or Sc atop an anionic O interlayer adsorbed to W\footnote{We adopt the notation that mixed adsorptions of A and B on C is notated A$_{X_A}$B$_{X_B}$/C, for $X_A$ the fraction of surface sites of C occupied by A. Adsorbed A-B complexes on C are notated A-B/C.} yield strong surface dipoles and enhanced thermionic emission. 
Given that the (001), (110), and (112) crystallographic planes are most commonly expressed in scandate cathodes (see, e.g. \cite{Liu:MC:2019}), a number of investigations of possible Ba$_X$Sc$_Y$/O configurations on these W facets have been made using first-principles calculations. 

First considering the W (001) surface, M{\"u}ller \cite{Muller:ASS:1997}, based on results of quantum mechanical calculations, proposed that Sc and O atoms occupy alternate 4-fold hollow sites on the W (001) surface while Ba atoms sit atop O atoms. 
Computationally evaluating model scandate cathode surfaces including (i) monolayer of Ba-Sc-O on W (100), (ii) Ba or BaO on Sc$_2$O$_3$+W, and (iii) BaO/(Sc$_2$O$_3$+WO$_3$), M{\"u}ller reported that the lowest work function was exhibited by a Ba-Sc-O monolayer on W (001) with a Ba coverage of about 1$\times$10$^{14}$ atoms/cm$^2$.
However, this proposed structure was reported to be neither thermodynamically stable, nor to possess a low work function by subsequent density functional theory (DFT) studies \cite{Vlahos:PRB:2010}.
A wide range of additional surface structures of Ba$_X$Sc$_Y$/O at different compositions, coverages, and adsorption sites on W (001) were studied by Vlahos et al. \cite{Vlahos:PRB:2010,Vlahos:APL:2009} using DFT.
These studies reported that O/W (001) surfaces that were partially covered with Ba exhibited a lower work function than those fully covered by Ba.
The lowest work function (1.16 eV) was achieved for a mixed Ba$_{0.25}$Sc$_{0.25}$/O/W (001) configuration, comparable to values reported for scandate cathodes \cite{Yamamoto:RPP:2006,Gartner:ASS:1997,Wang:ASS:2003a,Wang:ITED:2007}.

\begin{figure}
	\centering
	\includegraphics[width=1\linewidth]{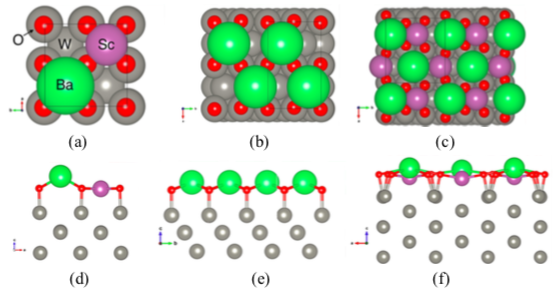} 
	\caption{top (a-c) and side (d-f) views of the positions of Ba, Sc, and O atoms for Ba$_{0.25}$O and Ba$_{0.25}$Sc$_{0.25}$O on a 2×2 W (100) surface (a, d), and Ba$_{0.25}$O on 2$\times$2 W (112), and Ba$_{1/3}$Sc$_{1/3}$O on W (112)\cite{Zhou:ASS:2018}} 
	\label{fig:surfacedecoration}
\end{figure}

Additional DFT studies\cite{Zhou:ASS:2018} using reference chemical potentials for Ba and Sc based on BaO and Sc$_2$O$_3$ and accounting for changes in O chemical potential also considered the Ba$_{0.25}$Sc$_{0.25}$/O/W (001) configuration.
In this configuration a full monolayer (ML) of O atoms are located directly atop surface W atoms, with Ba (0.25 ML) and Sc (0.25 ML) sitting above alternating hollow sites centered above four neighboring O atoms (Fig. \ref{fig:surfacedecoration}(a, d)).
These calculations reported that while Ba$_{0.25}$Sc$_{0.25}$/O/W (001) does exhibit the lowest work function (here computed to be 0.82 eV), Ba$_{0.25}$O on W (001) (with a work function of 1.23 eV) had a lower surface energy, and therefore would be expected to be observed in scandate cathodes in place of Ba$_{0.25}$Sc$_{0.25}$/O/W (001).
The stable Ba$_{0.25}$O on W (001) surface exhibited a low work function configuration (1.23 eV, similar to those reported for scandate cathodes\cite{Yamamoto:RPP:2006,Gartner:ASS:1997,Wang:ASS:2003a,Wang:ITED:2007}) and is structurally similar to Ba$_{0.25}$Sc$_{0.25}$/O/W (001), simply with the Sc sites vacant. 

While (001) surface configurations are the most widely studied, experimental evidence shows that (110) and (112) facets are also exposed on cathode surfaces \cite{Liu:MC:2019}.
Studies by Wan and Kordesch \cite{Wan:U:2012} and Webster et al. \cite{Webster::1963} show that electron emission varies strongly between different W crystal facets.
DFT calculations similar to those conducted to study BaSc/O/W (001) surface configurations have also been applied to (Ba or BaSc)/O/W (110) and (112) surface configurations \cite{Zhou:ASS:2018, Jacobs:AM:2017,Seif:ASS:2021}.

Zhou et al. \cite{Zhou:ASS:2018} reported that the lowest work function reported for any surface configuration on W (112) was 1.20 eV for Ba$_{0.50}$O/W (112), sufficiently low to be similar to work functions reported for scandate cathodes\cite{Yamamoto:RPP:2006,Gartner:ASS:1997,Wang:ASS:2003a,Wang:ITED:2007}.
Among W (112) surface configurations containing Sc, the lowest work function was slightly higher, at 1.45 eV for Ba$_{1/3}$Sc$_{1/3}$O/W (112).
Figure~\ref{fig:surfacedecoration} shows the structure of both Ba$_{0.50}$O/W (112) and Ba$_{1/3}$Sc$_{1/3}$O/W (112).
The relative stability of those surface configurations was reported to depend on the O chemical potential when the energies of Ba and Sc were referenced to BaO an Sc$_2$O$_3$. 
Based on observations of BaAl$_2$O$_4$, rather than BaO, in scandate cathodes \cite{Liu:MC:2019,Liu:M:2019}, combined with the possible presence of other Sc-containing compounds, Zhou et al.\cite{Zhou:ITED:2018} extended these results to also calculate energies of a number of surface configurations as a function of a full range of possible chemical potentials of Ba, Sc, and O. 
These extended calculations allowed comparison of surface energies of different surface configurations under any reasonable chemical conditions in which scandate cathodes might be found.
The relative stabilities of surface configurations were reported to vary significantly with different chemical conditions, but stable, low work function surfaces were often reported \emph{not} to include Sc, leading to the suggestion that the role of Sc was to modify the chemical environment around cathodes to stabilize low work function surfaces.

Some experimental studies have also focused on specific W facets.
Researchers at Osaka University \cite{Tsujita:SS:2003,Iida:SS:2003,Nagatomi:SS:2003} deposited on the order of 1 ML of Sc and O on W (001) surfaces, and reported that Sc-O complexes form at the W surface after heating. 
This was concluded based on AES peak position shifts for Sc and O compared to those of Sc-W  and Sc$_2$O$_3$, and indicated an Sc-O ratio of about 1:1. 
This result is consistent with the observed behavior of  Ba-O/W\cite{Iida:SS:2003,Nagatomi:SS:2003}. 
Also, a reversible (surface) phase transition was observed via low energy electron diffraction (LEED) for Sc-O/W (001) between a (1$\times$1) surface observed after heating at 1500-1700 K in UHV conditions and a (2$\times$1)-(1$\times$2) surface formed after exposure to O \cite{Nakanishi:SIA:2006,Iida:SS:2003,Nagatomi:SS:2003,Nakanishi:SS:2008,Tsujita:SS:2003,Nakanishi:SS:2005,Nakanishi:ASS:2009,Nagatomi:JVSTA:2010,Iida:SIA:2003}.

Using Auger Electron Spectroscopy (AES) and the secondary electron method (described in Ref.~\cite{Eyink:AoSS:1985}), these researchers studied the work function of ScO/W (001) surfaces \cite{Nakanishi:SS:2005,Nakanishi:SIA:2008,Iida:JSA:2002}.
They reported that a (1$\times$1) ScO/W (001) surface configuration with an ScO coverage of $\sim$0.6 ML resulted in a maximum reduction of 1.5--1.8 eV in work function compared to bare W (001). 
Other experimental studies have examined more generally the coverages, compositions, and atomic configurations of Ba, Sc and O on W that result in optimal scandate cathode performance.
Yamamoto et al. \cite{Yamamoto:JJAP:1988,Yamamoto:JJAP:1989,Yamamoto:V:1990} reported that a monoatomic layer of Ba-Sc-O reduces the work function of impregnated cathodes initially coated with W-Sc$_2$O$_3$.
This improvement in emission was hypothesized to originate with the oxidation of W-Sc$_2$O$_3$ into Sc$_2$W$_3$O$_{12}$, which reacts with Ba to form free Sc atoms.
The metallic Sc, along with Ba and O, are adsorbed to produce the 1 ML surface layer.
The formation of Sc$_2$W$_3$O$_{12}$ is required since cathode activation/operating temperatures are not high enough to decompose Sc$_2$O$_3$ directly \cite{Huber:JPC:1963,Reichl:ASS:1988}. 
Other studies by Sasaki et al. \cite{Sasaki:ASS:1999}, Wan and Kordesch \cite{Wan:JVSTBNMMPMP:2013}, and Liu et al. \cite{Liu:I25IVESCPICN:2004} support the idea that free Sc atoms at the surface supplied by chemical reactions inside the W matrix form a surface monolayer consisting of Ba, Sc, and O.
Additional discussion of plausible reactions for the production of free cations is included in the Appendix.

\subsubsection{Sc intermixing}
Other studies have suggested that intermixing of Sc (or Sc$_2$O$_3$) and W plays a role in achieving optimum emitting surface configurations.
Uda et al. \cite{Uda:ASS:1999} studied the composition of polycrystalline W with a 50 nm Sc$_2$O$_3$ overlayer both before and after heating using AES depth profile analysis.
They reported significant interdiffusion of Sc$_2$O$_3$ and W after heating for 3000 hours at temperatures of 1220 K, 1300 K, and 1370 K. 
It was hypothesized that Sc did not dissociate from O during diffusion as the ratio of Sc and O was always observed to be 1:1.5 as in Sc$_2$O$_3$.
The reported diffusion coefficients of Sc into W were 6.4$\times$10$^{-19}$, 1.0$\times$10$^{-18}$, and 1.6$\times$10$^{-18}$ cm$^2$/s at the temperatures studied. 
Shih et al. \cite{Shih:ASS:2005} reported observations of less BaO evaporation from Sc$_2$O$_3$ coated W surfaces than bare W, utilizing TPD and AES surface characterization techniques. 
Based on these results in conjunction with those of Uda et al., Shih et al. \cite{Shih:ASS:2005} posited that interdiffusion of Sc$_2$O$_3$ and W occurred during heating and Ba-O layers bond more strongly to the interdiffused layer than to pure W. 

Following up previous work \cite{Nakanishi:SS:2005} discussed above in which a 0.6 ML (1$\times$1) ScO/W (001) surface configuration exhibited a minimum work function, Nakanishi et al.\cite{Nakanishi:ASS:2009}, later reported that after exposure to O$_2$ at 5$\times$10$^{-7}$ Pa at 1500 K, the surface coverage of ScO decreased rapidly while coverage of Sc$_2$O$_3$ and work function increased rapidly.
This transition to coverage with Sc$_2$O$_3$ was accompanied by a surface phase transformation to a (2$\times$1)-(1$\times$2) surface configuration with a final work function comparable to bare W. 
This transition was explained as the result of diffusion of Sc-O complexes into the bulk W substrate and adsorption of O on the W (001) surface. 
The reverse transition was suggested to occur when Sc-O in the W substrate segregated to the surface after desorption of O from W (001), consistent with observed changes in work function \cite{Iida:SS:2003,Nagatomi:SS:2003,Nakanishi:SS:2008,Tsujita:SS:2003,Nakanishi:SS:2005,Nakanishi:ASS:2009}. 
Despite these findings, no interdiffused layer has been observed in high-resolution characterization of cross-sectioned scandate cathodes\cite{Liu:MC:2019}.

\subsubsection{Sc and cation desorption}
While the detailed configuration of adsorbates on W surfaces is widely acknowledged to be critical to determining cathode emission behavior, these configurations are not necessarily invariant over the life of a cathode.
In B-type cathodes the depletion of available Ba via evaporation from electron emitting surfaces has been understood to be a lifetime-limiting effect \cite{Yamamoto:RPP:2006,Cronin:IP:1981,Longo:1IEDM:1984,Longo:JAP:2003,Longo:1IEDM:1978,Rittner:JAP:1957,Roquais:MDiVES:2020,Roquais:ASS:2003}. 
It has also been reported that (near) steady-state surface coverages of Ba during operation of B-type and M-type cathodes were both less than optimal to achieve maximum work function reduction \cite{Cortenraad:ASS:2002}. 
Based on this perspective, stabilization of adsorbed Ba--hence increased Ba coverage (maximizing work function reduction) and/or cathode lifetime--has been proposed as a potential mechanism by which Sc enhances cathode performance.

Experimental support for this perspective can be found in the results of Zagwijn et al. \cite{Zagwijn:ASS:1997}, who studied a model system consisting of either Ba or BaSc deposited on W (100) in UHV conditions.
Both Ba and BaSc deposition was followed by exposure to oxygen at pressures of 5$\times$10$^{-9}$ mbar.
A surface exhibiting a low work function (as low as 1.18 eV) with a c(2$\times$2) LEED pattern was observed to be formed for all cases of BaSc deposition.
This surface was reported (via AES) to consist of $\sim$1 ML (each) of Ba, Sc, and O, similar to AES findings for scandate cathodes\cite{Hasker:ASS:1986,Hasker:ASS:1985}.
While the authors observed that Ba-only deposition could yield a lower work function surface at Ba coverages of $\sim$0.25 ML, this Ba coverage was not stable.
Low work function was stabilized by the addition of Sc, which was reported to manifest itself in an interlayer atop W on which Ba-O dipoles were adsorbed.
This ``Sc interlayer'' configuration differs from the BaSc/O/W layers usually considered with atomistic calculations (see, e.g. \cite{Zhou:ASS:2018,Vlahos:PRB:2010}).

Attention has also been given to whether Sc on W surfaces is stable or desorbs over time during cathode activation and operation. 
After reporting that Sc metal dewets from the W (100) at temperatures below 0.5$T_{\textrm{melt}}$\cite{Mroz:AA:2018, Mroz:JoVS&TB:2019,Mroz::2020,Mroz:JVSTB:2020}, Mroz et al. posited that the role of Sc in scandate cathodes leverages this desorption behavior to remove O from the W surface.
It was also reported that Sc$_2$O$_3$ reacts with Ba and/or BaO to form a compound that desorbs below cathode operating temperatures. 
Mroz et al. proposed that Sc metal does not interfere with work function reduction by Ba or with electron emission, rather enhancing these effects by cleaning the W of excess O \cite{Mroz:JoVS&TA:2019}. 

The potential for Sc to desorb from cathode surfaces during activation and operation has motivated efforts to enhance both Ba and Sc stability on cathode surfaces.
In one such approach, Lai et al. \cite{Lai:ASS:2018,Lai:ASS:2018a,Lai:ASS:2018b,Lai:JoAaC:2018} examined both Sc-free and Sc-doped cathodes fabricated with porous Re-W pellets. 
The addition of Re to W was reported to stabilize both Ba and Sc on cathode surfaces at elevated temperature.
As proposed by Mroz et al., Lai and co-authors support the idea that Sc desorbs from cathode surfaces in the form of Sc-O.
Lower diffusion barriers for the motion of Sc across Re (and Ru) surfaces were also reported, suggesting that surfaces containing these elements could more rapidly resupply Sc desorbed during activation and/or operation. 
Van Slooten and Duine \cite{VanSlooten:ASS:1997} also explored Sc cathodes with Re layers deposited onto the W pellet.
They used scanning Auger measurements to show that the resupply of Sc to the cathode surface dramatically slowed after the first few hours of cathode operation, suggesting that the presence of Re does not suppress the formation of highly stable Sc-containing compounds that reduce availability of free Sc.

Recently, a robust analysis of the chemical composition of material deposited on interior surfaces throughout a scandate cathode test vehicle after 30,000 hours of life testing was conducted by Seif et al. \cite{Seif:M:2020}.
XPS and EDS analysis did not reveal the presence of appreciable amounts of Sc on any test vehicle envelope or cathode assembly surface. 
This suggests that Sc did not desorb from the cathodes in any measurable quantity over the course of their 30,000+ hour lifetimes.

\subsubsection{Interplay of Sc and O}
Oxygen has been reported to play an important role in binding energy and work function reduction for alkali/alkali-metal adsorption on transition metals. 
It is well-accepted that the adsorbates at the surface of B-type cathodes are a Ba-O complex with Ba atoms as the topmost layer \cite{Cortenraad:ASS:2002,Cortenraad:JoAP:2001} and that the cathode reactivation rate relies on the resupply rate of O \cite{Cortenraad:JoAP:2001}. 
Work function reduction associated with increased O $2p$ binding energies was shown by Haas et al. \cite{Haas:AoSS:1983, Haas:ASS:1985} and was attributed to an enhanced dipole moment. 
A larger O $2p$ binding energy indicates that electrons are closer to O atoms, leaving more positive outward Ba atoms and resulting in an increased dipole moment \cite{Marrian:ASS:1985,Haas:ASS:1992}. 
Furthermore, the lowest work function for metal substrates are consistently reported as cases with both Ba and O adsorption, rather than \emph{only} Ba or O adsorbed surfaces \cite{Zhou:ASS:2018,Vlahos:PRB:2010,Vlahos:APL:2009}. 
Several studies reported the oxidization of Ba in scandate cathodes using AES peak analysis and found that Ba was partially metallic, similar to results from B-type cathodes \cite{Maloney:ASS:1985,Taguchi:ITED:1984}. 

Because of the importance of O in controlling work function, the role of Sc in regulating the availability of O at cathode surfaces has also been studied.
Hasker et al.\cite{Hasker:ASS:1985} were the first to propose that Sc directly influences O availability near cathode surfaces based on changes in surface composition observed with Auger microscopy. 
It was reported that there was an excess of O in B-type and M-type cathodes, but not in scandate cathodes. 
Oxygen concentration at the surface was even greater for degraded M-type cathodes, defined as having an emitting current decreased by a factor of $\sim$4 or more compared to initial post-activation emission. 
Although quantitative results for O concentration are limited by difficulties in determining sensitivity factors \cite{Hasker:ASS:1985}, they reveal clear trends in O concentration. 
It was reported that after activation, O concentration at the cathode surface is greater in scandate cathodes impregnated with BaO-CaO-Al$_2$O$_3$ in a ratio of 5:3:2 than in those with an impregnant ratio of 4:1:1. 
It also took a much longer time for 5:3:2 impregnant scandate cathodes to reach maximum emission than 4:1:1 cathodes -- 2000 hours vs. 2 hours, respectively. 
Furthermore, thermionic emission from scandate cathodes with (W+W/ScH$_2$)-top layers was reported to be higher than those with (W+Sc$_2$O$_3$)-top layers, which had a higher O concentration at the surface \cite{Hasker:ASS:1986}.

A more direct study on the influence of O partial pressure on cathode performance was reported by Maloney \cite{Maloney:ASS:1985}, where the effect of O$_2$ poisoning on the emission of scandate cathodes was investigated. 
These experiments were conducted by leaking O$_2$ into the test chamber with the ion getter pump off and then monitoring the O$_2$ partial pressure via an ionization gauge.
As the O$_2$ pressure P(O$_2$) increased, the emission current decreased only slightly at the outset, but a threshold P(O$_2$) value of about 1$\times$10$^{-4}$ Pa was reported, beyond which the emission decreased to almost zero. 
However, emission was reported to recover immediately as O$_2$ was depleted. 
The observation of high emission maintained before the threshold P(O$_2$) value indicated that the low work function surface configuration can be stable at a range of P(O$_2$) lower than this threshold. 
This is consistent with the work of Zhou et al.\cite{Zhou:ASS:2018,Zhou:ITED:2018}, who reported that low work-function surface configurations were only stable in certain oxygen chemical windows. 
Furthermore, this proposal was supported by the greater affinity of O to Sc than that to Ba. For the reaction:
\begin{equation}
	\text{3BaO + 2Sc $\leftrightharpoons$ Sc$_2$O$_3$ + 3Ba}
\end{equation}
$\Delta$G$^\circ$ = -26 kcal/(mol Ba) at 1200 K \cite{Hasker:ASS:1986,Barin::2013}. 
Similarly, Sc was reported to reduce WO$_2$ in the system with W, WO$_2$, and Sc$_2$O$_3$ by formation of Sc$_6$WO$_{12}$ \cite{Hasker:ASS:1986,Levitskii:IM:1981}. 
This may suggest why Sc, rather than Ba or W, is able to regulate O$_2$ partial pressure and stabilize low work function surface configurations.

\subsubsection{Effects of chemical environment}
While oxygen clearly plays an important role in controlling cathode surfaces, the chemical availability of all species in a cathode system influence which surface configurations are stable.
The fact that high-performing Sc-in-W cathodes exhibit a characteristic W particle shape has been used to deduce both the chemical environment present during cathode activation and operation and the sets of specific surface configurations stable at those conditions.
Figure~\ref{fig:Wwulffshape} highlights results of SEM characterization of high-performing scandate cathode surfaces that reveal this characteristic shape to be micron-scale W grains with (001), (110), and (112) facets.
Ba (or BaO) nanoparticles on the order of 10 nm in diameter decorate these crystal facets. 
Liu et al. \cite{Liu:MC:2019} concluded these BaO/Ba nanoparticles were condensed islands formed by the de-wetting (during cooling) of the (uniform, order 1 ML) adsorbate layer present at cathode surfaces during operation. 
As these shapes emerge during activation at high temperature and are stable during operation at temperatures of $\sim$1200 K over tens of thousands of hours, these W particle shapes were taken to be minimum energy (Wulff) shapes.
Minimum energy crystal shapes are dictated by the relative surface energies of the facets present, and relative surface energies are dictated by the specific surface configurations at each facet.

Which facets are stable--and therefore emerge during activation/operation--is a complex interplay of chemical environment and the detailed chemistry and configuration of possible surfaces.
This is supported by Szczepkowicz et al. who highlighted that W faceting is strongly influenced by a number of factors including O$_2$ partial pressure, temperature, and the presence of W dopants and/or surface adsorbates \cite{Szczepkowicz:SS:2011,Szczepkowicz:SS:2012}.
Given this, Zhou et al. \cite{Zhou:ASS:2018,Zhou:ITED:2018} used DFT to study 40 surface configurations of BaSc/O on the three different W crystal facets --- (001), (110), and (112) --- that terminate the characteristic W nanoparticles in scandate cathodes (Fig.~\ref{fig:Wwulffshape}).
Surface stability and minimum energy crystal shapes were calculated as a function of $\upmu_{\textrm{Ba}}$, $\upmu_{\textrm{Sc}}$, and $\upmu_\textrm{O}$, the chemical potentials---essentially a measure of chemical availability---of Ba, Sc, and O, respectively. 
Specific sets of surface configurations likely present in scandate cathodes were identified based on the criteria that the sets of W (001), W (110), W (112) surface configurations must both be stable (that is, exhibit low surface energies) and be able to yield the characteristic W shape observed in scandate cathodes (that is, exhibit specific relative surface energies).
Based on these criteria, two sets of surface configurations were identified as likely present in scandate cathodes: (i) Ba$_{0.25}$O/W (001), Ba$_{0.25}$O/W (110), and Ba$_{0.50}$O/W (112) and (ii) Ba$_{0.25}$Sc$_{0.25}$O/W (001), Ba$_{0.25}$O/W (110), and Ba$_{0.25}$O/W (112). 
The work functions computed for these sets of W (001), W (110), and W (112) surface configurations were consistent with those observed for scandate cathodes. 
The chemical environment necessary for these sets of surfaces to appear required similar Ba-rich and O-poor environments, but with set (i) requiring Sc-poor conditions and set (ii) Sc-rich. 
It was also reported that in extremely O-poor conditions Ba/W$_s$ [where $s$ represents (001), (110), and (112)] surface configurations were stable, and in O-rich conditions either O/W$_s$ or bulk WO$_3$ would form. 
Both of these cases lead to much higher work functions, implying that the chemical potential window in which the observed low work function W particle shapes are stable is highly constrained.
Moreover, Zhou et al. noted that while Sc is not necessarily present in the stable surface configuration, the low $\upmu_\textrm{O}$ (O-poor conditions) required to yield the observed particle shape requires the presence of Sc as an oxygen sink.

\begin{figure}
    \centering
    \includegraphics[width=1\linewidth]{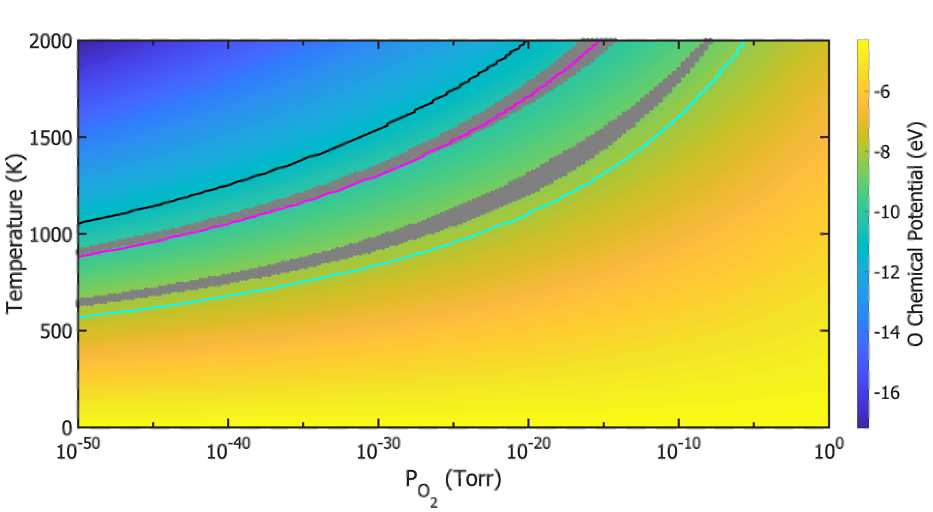}
    \caption{A ``phase diagram” relating $P_{\textrm{O$_2$}}$, $T$, and $\mu_\textrm{O}$. Gray points indicate the appearance of the characteristic W nanoparticle Wulff shape observed in high-performing scandate cathodes. Surface configurations including bare W, O/W, Ba/O/W, and BaSc/O/W were considered here.}
    \label{fig:seifdipolelines-contour}
\end{figure}

Seif et al. \cite{Seif:2I2ICVEI:2020a,Seif:2I2ICVEI:2022} used quantum mechanical calculations to directly explore the phase space of stable BaSc/O/W particle shapes accounting for effects of temperature and also the availability of all chemical species. 
This work reported computed surface excess free energies incorporating both temperature-dependent bulk and surface energy contributions.
These efforts incorporated both bulk and surface energy dependencies on temperature and accounted for bulk phase transformations (e.g., formation of BaO from Ba at moderate O availability) as a function of environmental conditions.
Finite temperature effects were incorporated by utilizing density functional perturbation theory (DFPT) to compute properties of phonons from DFT, including the phonon density of states and vibrational contributions to entropy, based on the approach of Togo et al. \cite{Togo:SM:2015}.
Seif et al. explored an array of surface configurations including bare W, O/W, Ba/O/W, and BaSc/O/W arrangements on (001), (110), and (112) facets.
It was reported that no surface configuration on any facet containing both Ba and Sc (e.g. BaSc/O/W) was the most stable at any considered temperature and oxygen availability.
However, the characteristic Wulff shape shown in Fig.~\ref{fig:Wwulffshape} was reported to be stable under certain conditions, but only when exhibiting surface terminations lacked Sc.
Overall the highest stability surfaces were either bare W (at extremely low O$_2$ availability) or Ba/O/W (at higher O$_2$ availability) regardless of Sc availability.
These results are consolidated in the phase diagram shown in Fig.~\ref{fig:seifdipolelines-contour}.

\section{Summary and Outlook}
Scores of papers representing decades of work from groups all over the world have been dedicated to the study of scandate cathodes.
We have attempted to capture a reasonable snapshot of both the breadth and depth of this work, ranging from fundamental to applied studies.
Overall it has been a goal of this review to try to capture the leading hypotheses proposed to explain the enhanced emission observed from Sc-containing cathodes.
In seeking this goal, though, we have highlighted that there is no consensus as to the role Sc plays in enhancing emission, nor as to the optimal fabrication processes, microstructure, or performance characteristics of scandate cathodes.

Some factors increasing the difficulty of revealing the processing-structure-properties relationships governing scandate cathodes are quite fundamental.
For example, the porous, rough, and (at least in some cases) coated emitting surfaces present in scandate cathodes cannot be assumed to function as a singular, idealized, semi-infinite flat surface.
Therefore it is not clear that a single-valued work function is applicable in exploring or reporting the performance of cathodes as devices.
Moreover, in applying the Richardson-Dushman equation to extract ``effective'' work functions for cathodes, the assumption of a Richardson constant of 120 A/cm$^2$K$^2$ should certainly be questioned.
Finally, it is far from clear which near-Fermi level states are contributing electrons to the (thermionic) emitted current.
For example, are emitted electrons liberated from surface states unique to particular facets or surface features, or from bulk states characteristic of the supporting W particles (and whose emission is then modified by surface-related conditions)?

Turning to microstructure, we have discussed the most common fabrication approaches--here distinguished as \emph{Sc-in-W, Sc-in-impregnant, Sc-in-oxides,} and \emph{top layer} approaches-- for scandate cathodes and presented the microstructures and emission behaviors observed for each.
While the fabrication approaches differ with respect to details of precursor powders, impregnant composition, and the timing and method of Sc introduction, available imaging results reveal microstructures that share common features: a highly porous emitting material consisting of W particles decorated with Ba, Sc, and O.
Despite this, the paucity of comprehensive and detailed characterization results available in the published literature--though perhaps due to concerns about proprietary processes or materials--represents a major hurdle for efforts aiming to uncover and leverage knowledge of processing-structure-properties relationships.

Considering cathode performance, observed current densities reported for scandate cathodes fabricated with different methods generally fall in a range.
These reports agree that scandate cathodes yield 1-2 orders of magnitude higher current densities at $\sim$100 K lower operating temperatures compared to previous non-scandate cathodes.
However, most literature reports citing emission data lack detailed descriptions of the cathodes tested, the mode of cathode operation, measurement conditions, and the configuration of device components during testing.
As such, assessment and comparison of reported emission properties is challenging.

Lastly we have explored three classes of mechanisms proposed to explain how Sc enhances electron emission: the \emph{Ba adsorbed on Sc-containing layer} model, the \emph{semiconductor} model, and the \emph{dipole} model. 
Each of these models originated from studies of previous generations of thermionic cathodes, rather than having been formulated specifically for Sc cathodes.
The \emph{Ba adsorbed on Sc-containing layer} model posits that the emitting surface is Ba/BaO adsorbed on a bulk-like Sc-containing layer, while the \emph{semiconductor} model presupposes the existence of a similarly bulk-like semiconducting coating at the cathode surface.
To date though, detailed imaging of scandate cathodes (via, e.g., SEM) with resolution on the nanometer scale does not reveal the presence of such layers.

Even the \emph{dipole model}, which does not invoke the presence of a thick or bulk-like surface layer, is challenging to directly apply to scandate cathodes.
As noted above, reports in the literature variously suggest that Sc alters chemical conditions leading to the formation of surfaces with optimal dipole configurations, is present on cathode surfaces directly enhancing surface dipole strength, retards desorption of cations from the surface at operating conditions (thereby enhancing the dipole effect on work function), or getters O (anions) from the surrounding chemical environment.
It is wholly possible that all, some combination of, or none of these mechanisms ultimately explain the role of Sc in enhancing emission from scandate cathodes.
Reports available in the literature to date do not provide sufficient basis for a clear conclusion.

Looking forward to continued improvement in understanding, performance, and control of scandate cathodes, a set of specific research needs stand out.
Given the numerous and complex steps in cathode fabrication, rigorous and complete reporting of processing and activation conditions is critical for linking processing and performance.
In aggregate, the available literature suggests that minor variations in fabrication details (e.g. activation temperature/time, presence or absence of applied fields and/or vacuum/environmental conditions during fabrication, among many other factors) can have significant effects on resulting cathode properties.
This need is particularly acute when attempting to link controlled studies on model systems--which are essential for isolating and demonstrating specific mechanisms and relationships--to more applied studies of full cathodes.
In many cases the potential impact of otherwise excellent studies is diminished because, in comparing or connecting to previous studies, it is unclear whether fabrication and/or processing conditions were equivalent.

In addition, comprehensive characterization campaigns are needed to reveal the sub-micron structure and composition of cathodes resulting from different fabrication approaches.
Specifically it is necessary to pair imaging techniques (e.g. via SEM and/or TEM) with composition sensitive non-imaging techniques (e.g. AES, EDS).
Approaches for characterizing cathode surfaces and cathode cross-sections at all stages of cathode fabrication must also be prioritized.
The composition of species desorbed during operation (measured e.g. via RGA) would provide insight into the ``dispenser" nature of Sc-containing cathodes.

Lastly, there is a significant need for a standardized approach for measuring and reporting thermionic emission from cathodes and cathode materials.
Optimizing cathodes as devices requires individual knowledge of how emitting materials themselves behave under operating conditions.
A measurement approach based on close-spaced diode (CSD) testing may be appropriate to address this need, but care must be taken with respect to the design and configuration of test apparatus to ensure consistent control of current leakage, temperature (and temperature measurement), and current collection.

Researchers from a broad spectrum of the scientific community have contributed and continue to contribute to the study of thermionic cathodes generally and scandate cathodes specifically.
The efforts of this community have undoubtedly made significant contributions to understanding and optimizing Sc-containing cathodes, and powerful tools and capabilities have been developed for investigating cathodes.
Despite this, little consensus has emerged as to the composition, microstructure, and/or fabrication processes that optimize Sc-containing cathodes, and few, if any, applied systems currently employ Sc-containing cathodes.
In seeking to address this, the complexity of Sc-containing cathodes makes clear that developing sufficient understanding and control of these devices to allow their reliable integration into commercial or military applications will require a dedicated effort to effectively and comprehensively communicate experimental details and results among researchers with disparate backgrounds and perspectives.

\section{Acknowledgments}
This work was financially supported by the Defense Advanced Research Projects Agency (DARPA) Innovative Vacuum Electronics Science and Technology (INVEST) program, under grant number N66001-16-1- 4041. The views, opinions, and/or findings expressed are those of the author(s) and should not be interpreted as representing the official views or policies of the Department of Defense or the U.S. Government.

Thanks to, and in memory of, Dr. Eric A. Grulke for many helpful comments and a lifetime of research, teaching, and service.

\section{Appendix}
\subsection{Formation and Transport Mechanisms of Free Metallic Ba and Sc}
For B- and M-type cathodes, a key process occurring during cathode activation and operation has been understood to be the generation of free Ba atoms (or BaO complexes) available to migrate to the cathode surface.
Generally the liberation of these species from immobile, high melting temperature oxides has been understood to result from reactions between impregnants and the W/W-alloy matrix \cite{Yamamoto:RPP:2006,Cortenraad:ASS:2002,Rittner:JAP:1957}.
Several reactions for the production of atomic Ba have been proposed. 
In 1957, Rutledge and Rittner \cite{Rutledge:JAP:1957} studied L-type cathodes, which were comprised of a porous W matrix covering a pocket of Ba compound containing BaO or mixed BaO and SrO. 
It was proposed that free Ba--that is, Ba not confined to a chemical compound in the Ba reservoir--was generated by the reaction:
\begin{equation}
	\text{2BaO + $\frac{1}{3}$W $\rightarrow$ $\frac{1}{3}$Ba$_3$WO$_6$ + Ba}
\end{equation}
The equilibrium Ba partial pressure associated with the evaporation of Ba evolved from this reaction at elevated temperature was found to be well-represented as:
\begin{equation}
	\log \text{P$_{\textrm{Ba}}$ (mmHg) = -16400/$T$ + 8.02} \label{ba_evap} 
\end{equation}
 
X-ray diffraction analysis of constituents on the bottom of the W pellet at the end of the operational life of the cathode identified significant amounts of Ba$_3$WO$_6$, supporting this proposed mechanism. 
Subsequent heating of the failed cathode did result in further evaporation of atomic Ba, but the evaporation rate decreased with time while the rate of evaporating BaO increased.
During this post-lifetime heating no thermionic emission was detected.
Based on these findings, a second stage of Ba generation was proposed to occur via\cite{Rutledge:JAP:1957}:
\begin{equation}
	\text{$\frac{2}{3}$Ba$_3$WO$_6$ + $\frac{1}{3}$W $\rightarrow$ BaWO$_4$ + Ba} \label{free_Ba}
\end{equation}
This was again supported by X-ray diffraction, which found BaWO$_4$ to be present after post-lifetime heating.
BaWO$_4$ was found to be highly stable, implying that once formed no additional free Ba could be produced to resupply Ba desorbed from the cathode surface.
Similar results were found for cathodes where BaO replaced Ba$_3$WO$_6$ as the initial (first stage) Ba source. 
These findings led to a general understanding that an increase in the presence of BaWO$_4$ (that is, an increase in the relative O concentration in Ba containing oxides) was associated with the observed end-of-life degradation in cathode performance, as less and less free Ba remained available at the surface.

Similar mechanisms were proposed for cathodes impregnated with Ba aluminates. 
XRD of cathodes with Al-containing impregnants revealed the presence of a mixture of Ba$_3$Al$_2$O$_6$ and BaAl$_2$O$_4$ after impregnation with a nominal composition of 5BaO-2Al$_2$O$_3$ \cite{Rittner:JAP:1957}.
For these cathodes, it was proposed by Rittner et al.\cite{Rittner:JAP:1957} that free Ba was generated via:
\begin{equation}
	\text{$\frac{2}{3}$Ba$_3$Al$_2$O$_6$ + $\frac{1}{3}$W} \rightarrow
	\text{$\frac{1}{3}$BaWO$_4$ + $\frac{2}{3}$ BaAl$_2$O$_4$ + Ba} 
\end{equation}
In this case, the equilibrium Ba partial pressure was found to vary as:
\begin{equation}
\log \text{P$_{\textrm{Ba}}$ (mmHg)} = -20360/T + 8.56 \label{Rittner_equil} 
\end{equation}
For BaO+Al$_2$O$_3$ impregnants, a related reaction was proposed \cite{Rittner:JAP:1957}:
\begin{equation}
	\text{5BaO + 2Al$_2$O$_3$ $\rightarrow$ $\frac{3}{2}$Ba$_3$Al$_2$O$_6$ + $\frac{1}{2}$BaAl$_2$O$_4$}
\end{equation}
with Ba$_3$Al$_2$O$_6$ decomposing to release Ba as above.
BaAl$_2$O$_4$ was found to be chemically inert, similar to BaWO$_4$, and, intriguingly consistent with findings 60 years later by Liu et al. \cite{Liu:MC:2019} who observed BaAl$_2$O$_4$ nanoparticles decorating the surfaces of scandate cathodes after extended operation.

Similarly, generation of free Ba in BaO-CaO-Al$_2$O$_3$ (in a ratio of 5:3:2) mixed impregnants was proposed to occur via Rittner et al. \cite{Rittner:JAP:1977}:
\begin{multline}
	\text{5BaO$\cdot$3CaO$\cdot$2Al$_2$O$_3$ + $\frac{3}{4}$W} \rightarrow \\
	\text{$\frac{3}{4}$Ca$_2$BaWO$_6$ + $\frac{3}{4}$Ca$_2$BaAl$_2$O$_6$ + $\frac{5}{4}$BaAl$_2$O$_4$ + $\frac{9}{4}$Ba}
\end{multline}
A related alternative reaction for the generation of free Ba from a similar impregnant mixture was suggested by Cronin \cite{Cronin:IP:1981}:

\begin{multline}
 	\text{W + 3Ba$_3$Al$_2$O$_6$ + 6CaO}  \rightarrow \\
	\text{3Ba$_2$CaAl$_2$O$_6$ + Ca$_3$WO$_6$ + 3Ba}
\end{multline}
where Ba$_3$Al$_2$O$_6$ was formed from BaO-Al$_2$O$_3$ as shown above.

Yamamoto et al. \cite{Yamamoto:RPP:2006} proposed a reaction that also generated free Ca (in addition to free Ba) from BaO-CaO-Al$_2$O$_3$ impregnants:

\begin{multline}
	\text{5BaO$\cdot$3CaO$\cdot$2Al$_2$O$_3$ + W$\rightarrow$} \\
	\text{2BaAl$_2$O$_4$+$\frac{3}{4}$ Ca$_2$BaWO$_6$}+ \\ \text{$\frac{1}{4}$Ca$_3$WO$_6$ + $\frac{9}{4}$Ba + $\frac{3}{4}$Ca}
\end{multline}

Efforts have been made to apply the above approach to scandate cathodes, focusing on the hypothesis that free Sc plays a similar role to free Ba in generating surface dipoles\cite{Vaughn:ITED:2009}.
Yamamoto et al. \cite{Yamamoto:ASS:2005,Yamamoto:JJAP:1989,Yamamoto:JJAP:1988} proposed that free Ba participated in generation of free Sc according to the reactions:
\begin{equation}
	\text{Sc$_2$O$_3$ + 3WO$_3$} \rightarrow \text{Sc$_2$W$_3$O$_{12}$}
\end{equation}
and
\begin{equation}
	\text{Sc$_2$W$_3$O$_{12}$ + 3Ba $\rightarrow$ 3BaWO$_4$ + 2Sc}
\end{equation}
This was proposed based on study of scandate cathodes with top-layer coatings consisting of W+Sc$_2$O$_3$ \cite{Yamamoto:JJAP:1988} or W+Sc$_2$W$_3$O$_{12}$ \cite{Yamamoto:JJAP:1989}, where it was concluded that only free Ba was available to react with the Sc-containing coatings as neither Al$_2$O$_3$ nor CaO were detected at the cathode surface.

Liu et al. \cite{Liu:ASS:2005} studied the change in Sc concentration at cathode surfaces during activation and reported an increase over time for impregnated cathodes but not for unimpregnated cathodes. 
The authors also reported that, directly prior to activation, no Sc was detected at the surface.
These findings indicated that Sc$_2$O$_3$ present in cathodes prior to activation does not migrate to the surface on its own, while the presence of impregnant results in a chemical reaction that liberates Sc, freeing it to migrate to the surface.
The absence of Sc at the cathode surface after washing was proposed to be a consequence of the formation of soluble Ba$_2$ScAl$_2$O$_5$ during impregnation:
\begin{equation}
	\text{4BaO$\cdot$CaO$\cdot$Al$_2$O$_3$} + \text{Sc$_2$O$_3$ $\rightarrow$} 
	 \text{ 2Ba$_2$ScAl$_2$O$_5$ + CaO} 
\end{equation}
This was supported by the detection of Ba$_2$ScAlO$_5$ using XRD after impregnation \cite{Liu:ASS:2005}. 

As the above reactions rely on impregnant material in the bulk of a cathode, a mechanism for transporting free Ba, Ca, and/or Sc to the exposed cathode surface is required.
This transport is expected to take one of two forms: (i) surface diffusion along internal pore surfaces or (ii) vapor flow through pores. 
Rutledge and Rittner \cite{Rutledge:JAP:1957} measured the flow rates of inert gases through W pellets of various porosities and found them to be consistent with observed Ba evaporation rates. 
Therefore, the Ba transportation mechanism was suggested to be Ba vapor flow through the pores as Knudsen flow \cite{Rittner:JAP:1957,Rutledge:JAP:1957}. 
However, while the melting point of Ba is very low (1000 K or 727$^\circ$C), its boiling point is 2080 K ($\sim$1800$^\circ$C), far exceeding that of the activation or operating temperature of B-type or scandate cathodes--implying that direct evaporation of bulk Ba is not the source of Ba vapor. 
In the case of Sc, Wang et al.\cite{Wang:ITED:2007} used \emph{in situ} AES to detect arrival of Sc at the surface starting at about 1000$^\circ$C$_b$ and Ba at about 800$^\circ$C$_b$, although mechanisms for Sc transport remain an area of community interest.

\bibliography{bibliography}
\bibliographystyle{unsrt}

\end{document}